\documentclass[aps,prd,twocolumn]{revtex4-2}

\usepackage[english]{babel}
\usepackage[utf8]{inputenc}
\usepackage[colorinlistoftodos, color=green!40, prependcaption]{todonotes}
\usepackage{amsthm}
\usepackage{mathtools}
\usepackage{physics}
\usepackage{xcolor}
\usepackage{graphicx}
\usepackage[left=23mm,right=13mm,top=35mm,columnsep=15pt]{geometry} 
\usepackage{adjustbox}
\usepackage{placeins}
\usepackage[T1]{fontenc}
\usepackage{csquotes}
\usepackage[pdftex, pdftitle={Article}, pdfauthor={Author}]{hyperref} 
\usepackage{setspace}
\usepackage{float}
\usepackage{graphicx}
\usepackage{subcaption}
\usepackage{booktabs}  
\usepackage{orcidlink}

\newcommand{\ttbar}{t\bar{t}}
\newcommand{\gev}{GeV}
\newcommand{\MG}{\textsc{MadGraph5\_aMC@NLO}}
\newcommand{\Pythia}{\textsc{Pythia8}}
\newcommand{\RF}{\textsc{RestFrames}}

\newcommand{\FastJet}{\textsc{FastJet}}
\newcommand{\MadAnalysis}{\textsc{MadAnalysis5}}
\newcommand{\MS}{\textsc{MadSpin}}
\newcommand{\ATLAS}{\textsc{ATLAS}}
\newcommand{\CMS}{\textsc{CMS}}
\newcommand{\chel}{c_{\rm hel}}
\newcommand{\nchel}{Nc_{\rm hel}}
\newcommand{\chan}{c_{\rm han}}

\begin{document}
\title{Analysing Toponium at the LHC using Recursive Jigsaw Reconstruction}

%
%

\author{Aman Desai \orcidlink{0000-0003-2631-9696}}
    \email{aman.desai@adelaide.edu.au}
    \affiliation{Department of Physics, The University of Adelaide, North Terrace, Adelaide, SA 5005, Australia}
\author{Amelia Lovison \orcidlink{0009-0009-5196-4647}}
    \email{amelia.lovison@adelaide.edu.au}
    \affiliation{Department of Physics, The University of Adelaide, North Terrace, Adelaide, SA 5005, Australia}
\author{Paul Jackson \orcidlink{0000-0002-0847-402X}}
    \email{p.jackson@adelaide.edu.au}
    \affiliation{Department of Physics, The University of Adelaide, North Terrace, Adelaide, SA 5005, Australia}

\date{\today} 
\onehalfspacing

\begin{abstract}
Recent results from the \ATLAS ~and the \CMS ~experiments at the Large Hadron Collider indicate the presence of a top-quark pair bound state near the  threshold region. We present a way to reconstruct a toponium state at the $t\bar{t}$ threshold region formed at the Large Hadron Collider using the Recursive Jigsaw Reconstruction. We have considered the Non-Relativistic QCD based toponium model implemented in MadGraph5\_aMC@NLO. The final states considered consist of two b-jets, two oppositely charged leptons, and missing energy that arises from two neutrinos. The goal of the Recursive Jigsaw Reconstruction is to make use of rules that can help resolve combinatorics ambiguity in preparing the decay tree for a given physics event. Additionally, missing energy coming from two neutrinos needs to be resolved in order to reconstruct the event. We apply four different methods within the \RF~package and compare the reconstruction resulting from each of the methods. Due to this method, one can also access kinematic variables in the rest frames belonging to intermediate particle states, providing additional means to discriminate the SM $\ttbar$ background from the toponium signal. We propose using two angular variables to enhance sensitivity to the toponium signal. Our preliminary results indicate that the improvement in sensitivity can be as much as 12\% over the current strategy in the LHC's Run 3 configuration. This method may be useful for gaining additional insight into the physics phenomenology in the $\ttbar$ threshold region.
\end{abstract}


\maketitle

\section{Introduction}

The ATLAS and the CMS experiments at the Large Hadron Collider have recently reported an excess of events near the top-quark pair production threshold in the dilepton decay channel~\cite{CMS:2025dzq,CMS:2025kzt,ATLAS:2026dbe}. The results are consistent with the presence of a spin-0 pseudoscalar, toponium~\cite{Fadin:1987wz,Fadin:1987wz,Hagiwara:2008df}. 

The reconstruction of this dileptonic decay mode of the $\ttbar$ system is complex as it involves the presence of two neutrinos which escape detection, and their presence is only inferred through indirect methods such as the imbalance of the transverse momentum. Additionally, owing to the nature of proton-proton collisions, the kinematic information of the neutrino's four-momentum along the beam axis is lost as the initial energy of a parton interaction is unknown. This leads to challenges in the reconstruction of the $\ttbar$ system.

The problem is mainly two-fold. First, as mentioned before, there are two neutrinos and, second, there are two b-jets, which leads to kinematic as well as combinatoric ambiguities. Several methods relying on analytical or geometrical techniques have been proposed. For instance, the Ellipse Method~\cite{Betchart:2013nba} and the Sonnenschein Method~\cite{Sonnenschein:2005ed,Sonnenschein:2006ud} employ analytical and geometrical solutions, which have been proposed to reconstruct the top-quark pair in the dilepton final state. The ATLAS experiment used the Ellipse Method to reconstruct the dilepton channel in their analysis~\cite{ATLAS:2026dbe}, while the CMS experiment used the Sonnenschein Method~\cite{CMS:2025kzt}.

In this paper, the Recursive Jigsaw Reconstruction method~\cite{Jackson:2017gcy} is used, which is based on the use of jigsaw rules or algorithms to deal with known and unknown combinatorics. We propose this as a new strategy to discriminate the toponium signal sample from the baseline $\ttbar$. Such a strategy might be useful in further understanding the physics of this quasi-bound state.

The paper is organised as follows: In Section~\ref{sec:mcstrat}, we present the details regarding the Monte Carlo event samples that we generated for the analysis and also list the pre-selection criteria that we have applied on the events. In Section~\ref{sec:recostrat}, we give an outline of the reconstruction strategy that we have followed to reconstruct the $\ttbar$ final state. In Section~\ref{sec:preanalysis}, we present the new variables and perform an analysis comparing the current strategy with the new analysis strategy using pre-selection criteria. Finally, in Section~\ref{sec:final}, we present the analysis applying a set of event selection criteria. 

\section{Monte Carlo Event Generation and Pre-selection}\label{sec:mcstrat}

In this study samples were generated according to the LHC Run 2 and Run 3 configurations. In particular, the simulations for Run 3 (Run 2) proton-proton collisions were carried out at $\sqrt{s} = 13.6$ TeV ($\sqrt{s} = 13$ TeV) centre-of-mass energies and the scaling with integrated luminosity uses $\mathcal{L} = 300  ~\rm{ fb}^{-1}$ for LHC Run 3 and  $\mathcal{L} = 140  ~\rm{ fb}^{-1}$ for LHC Run 2.

The simulation of the toponium samples follows the prescription outlined in Ref~\cite{Fuks:2024yjj} which incorporates matrix element re-weighting through Non-Relativistic Quantum Chromo Dynamics (NRQCD) Green's function. The procedure is implemented in \MG~and with the CT18NLO Parton Distribution Function set (PDF), obtained using LHAPDF6~\cite{Buckley:2014ana} library. This procedure allows us to simulate the toponium bound state ($\eta_t$) near the $\ttbar$ threshold region in the dilepton decay mode without introducing any intermediate particle. The Hard Scattered matrix elements simulated in \MG~\cite{Alwall:2014hca} are passed to \Pythia~\cite{Bierlich:2022pfr} for parton showering and hadronisation. We generated 1,000,000 events at $\sqrt{s} = 13$ TeV and 2,000,000 events at $\sqrt{s} = 13.6$ TeV centre-of-mass (CoM) energy. 

The background $\ttbar$~sample was generated using \MG~at Next-to-Leading Order (NLO) accuracy in QCD. This was followed by \MS~\cite{Artoisenet:2012st} to decay unstable, heavy intermediate particles. The events were then processed through \Pythia~for parton showering and hadronisation. Here the fixed-order matrix elements at NLO are matched with a parton shower as per the procedure in Ref~\cite{Alwall:2014hca}. The NNPDF3.0 PDF set~\cite{NNPDF:2014otw} was used for the Run 2 sample, and PDF4LHC21 PDF set~\cite{PDF4LHCWorkingGroup:2022cjn} was used for the Run 3 sample. In this study, the top quark mass is set to 173 GeV and the top quark width is set to $\Gamma_t=1.49$ GeV.  \MS~\cite{Artoisenet:2012st} was used to handle the decay of a $\ttbar$ pair to $bbW(\ell\nu)W(\ell\nu)$. 

The samples are then processed through \FastJet~\cite{Cacciari:2011ma}, accessed via \MadAnalysis~\cite{Conte:2018vmg,Araz:2020lnp} framework, to reconstruct jets using the anti-$k_T$ algorithm~\cite{Cacciari:2008gp} with a radius parameter of $R=0.4$. In the following study, the Run 3 $\ttbar$ baseline sample is normalised to $\sigma_{\ttbar}\times Br(W\rightarrow l\nu)^2=43.6$ pb  considering the NNLO+NNLL predictions~\cite{ATLAS:2023slx,CMS:2023qyl}, while the toponium sample is normalised to $\sigma_{\eta}\times Br(W\rightarrow l\nu)^2=0.277$ pb. The events are then weighted to the expected LHC Run 3 integrated luminosity of $300~\rm{fb}^{-1}$. For the Run 2 configuration we have normalised the samples using $\sqrt{s}=13$ TeV predictions- $\sigma_{\ttbar}\times Br(W\rightarrow l\nu)^2=36.9$ pb~\cite{Czakon:2013goa} and $\sigma_{\eta}\times Br(W\rightarrow l\nu)^2=0.25$ pb with the integrated luminosity set to $140~\rm{fb}^{-1}$. Here, the $\eta_t$ production cross-section was obtained via \MG~at Leading Order accuracy in QCD.

The events consisting of at least two oppositely charged leptons $(ee, \mu\mu, e\mu)$ and at least two b-quark initiated jets ($b$-jets) are selected. The leptons and $b$-jets are required to satisfy the conditions: $p_T^{\ell} > 25$ \gev, $|\eta|^{\ell} < 2.5$, $p_T^{b-jet} > 25$ \gev, $|\eta|^{b-jet} < 2.5$.  Moreover, a requirement on the angular distance $\Delta R(b\rm{-jet}, \ell) > 0.4$ is imposed. The pre-selection criteria are summarised in Table~\ref{tab:preselection}.

Unless stated otherwise, all of the plots, results and discussion in the following text are concerning the LHC Run 3 luminosity and energy conditions.  

\begin{figure}[!htbp]
    \centering
    \includegraphics[width=0.9\linewidth]{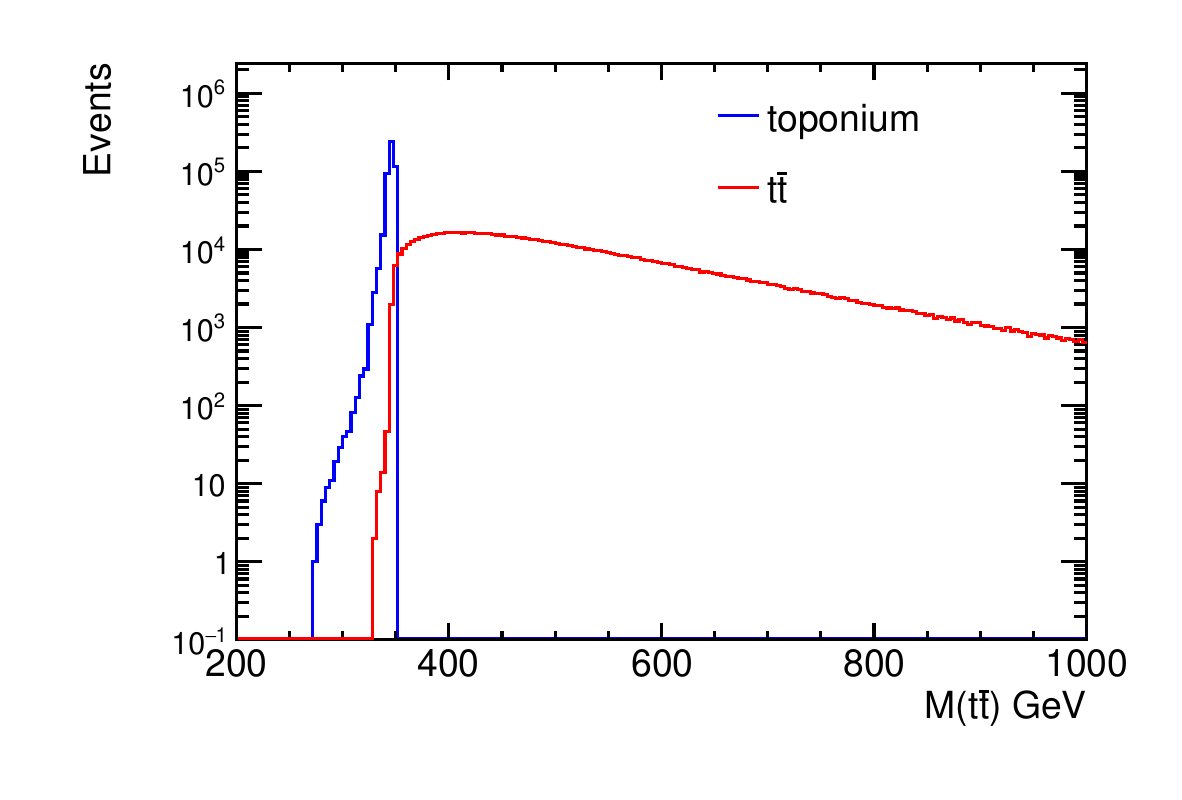}
    \caption{The $M_{\ttbar}$ mass at the truth level after applying pre-selection criteria.}
    \label{fig:mtttruth}
\end{figure}

The main differences between the $\eta_t$ sample and baseline $\ttbar$~sample is that the former was mediated through a pseudo-scalar quasi-bound state, whereas the latter process proceeds via spin-1 gluon mediation or through t-channel top-quark exchange. Moreover, the toponium sample exists only in the narrow range of $M_{\ttbar} \approx (330,350)$ GeV as seen in the truth distribution, Figure~\ref{fig:mtttruth}.

\begin{table}[!htbp]
\centering
\caption{Pre-selection criteria used in this analysis.}
\begin{tabular}{ll}
\hline\hline
\textbf{Selection} & \textbf{Requirement} \\
\hline
b-jet  multiplicity  & $N_{b-jet} \ge 2$ \\[2mm]
Lepton multiplicity  & $N_{\ell} = 2$ with $\Sigma_{\ell} N_Q = 0$ \\[2mm]
b-jet $p_T$ & $p_T^{b-jet} > 25$ \gev\\[2mm]
b-jet $\eta$ & $|\eta|^{b-jet} < 2.5$   \\[2mm]
Lepton $p_T$ & $p_T^{\ell} > 25$ \gev\\[2mm]
Lepton $\eta$ & $|\eta|^{\ell} < 2.5$   \\[2mm]
$\Delta R$($b$-jet,Lepton) & $\Delta R(b\rm{-jet}, \ell) > 0.4$ \\[2mm]
\hline\hline
\end{tabular}
\label{tab:preselection}
\end{table}

\section{Reconstruction Strategy for $\ttbar$ in dilepton final states}\label{sec:recostrat}

In order to reconstruct $\ttbar \rightarrow bbW(l\nu)W(l\nu)$ processes, which is a combinatoric challenge with both known ($b-jet$ pairing with lepton) as well as unmeasured particles (two neutrinos), we employ the Recursive Jigsaw Reconstruction method~\cite{Jackson:2017gcy}. As an overview, the algorithm functions by first choosing a decay tree that considers all intermediate decaying particles with measured and invisible final state particles. For each step in the decay tree, the algorithm applies a jigsaw rule that relates the current frame of reference to the next one in the chain as a function of measured and unknown quantities. Finally, these steps are applied recursively until the end of the decay chain. 

\begin{figure}[!htbp]
    \centering
    \includegraphics[width=0.7\linewidth]{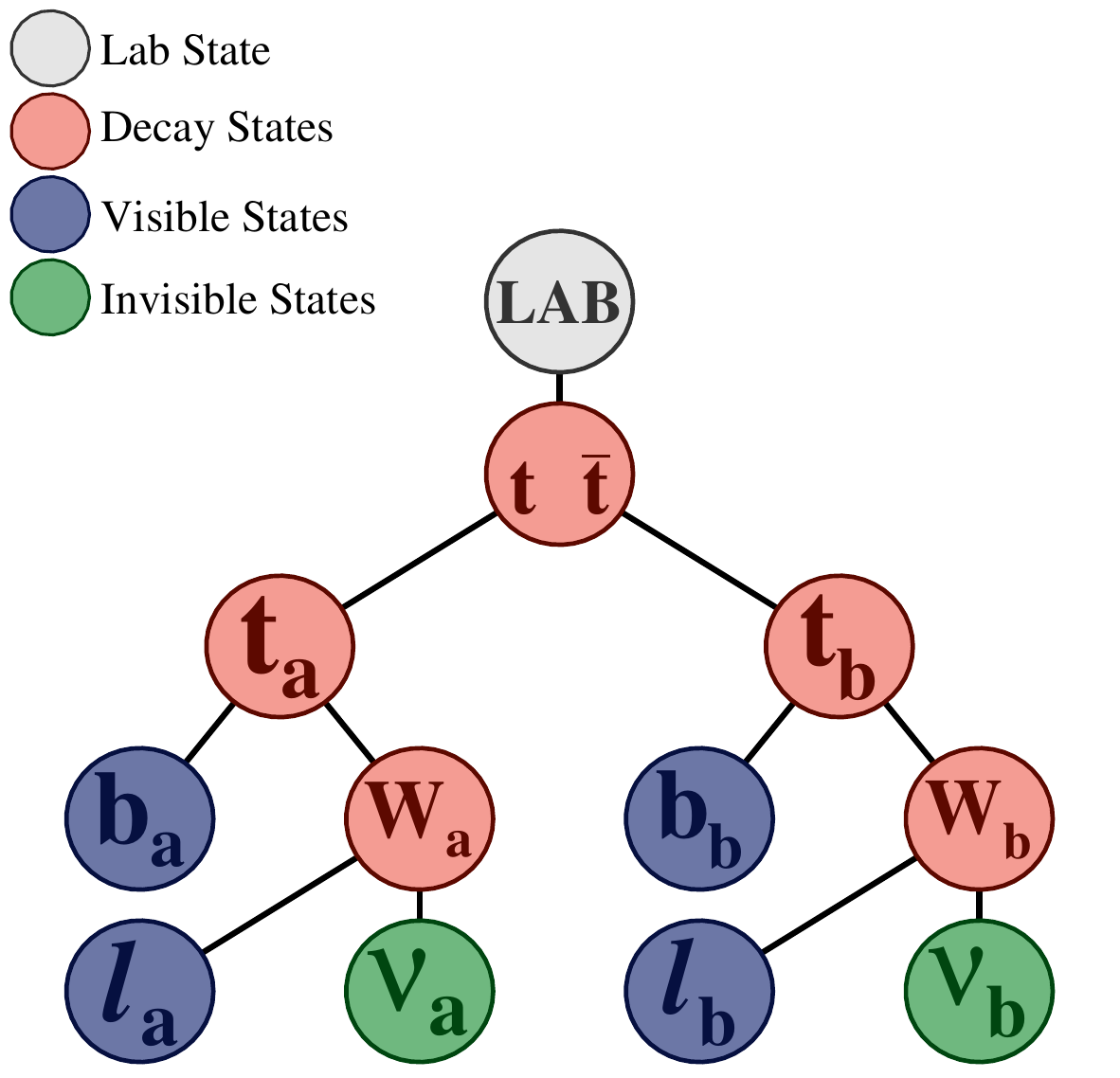}
    \caption{A decay tree diagram for the process $\ttbar\rightarrow bW(\ell\nu)bW(\ell\nu)$.}
    \label{fig:decaytree}
\end{figure}

The prescriptions proposed in~\cite{Jackson:2017gcy} for the $\ttbar$ dilepton channel is followed. Here, final states consisting of two b-jets and oppositely charged leptons which could be $ee$, $\mu\mu$, or $e\mu$ are considered. The missing transverse energy is considered as the transverse momentum of a dineutrino system. The decay tree used in the analysis is shown in Figure~\ref{fig:decaytree}. 

As shown in the decay tree, there are four visible final states and two invisible states. The algorithm first pairs the lepton and b-quark by minimising the following function:

\begin{equation}
    f = \rm{min} \{~M_{\ell_a, b_{a}}^2 + M_{\ell_b, b_{b}}^2\}
\end{equation}

where $M_{\ell_a, b_{a}}$ is the invariant mass of the pair of lepton and $b$-jet. The subscripts $a$ and $b$ refer to the ordering as indicated in \autoref{fig:decaytree}.

This is motivated as the combination of four four-vectors for decay products and is expected to be smaller for particles coming from a common parent particle. Since the final state consists of four visible particles originating at different decay steps, it allows one to choose how the total invisible system is split into two final state invisible particles. Namely, one can choose (i) $M(\ell_a, b_a, \nu_a) = M(\ell_b, b_b, \nu_b)$, which leads to the constraint $M_{ta} = M_{tb}$, or (ii) $M(\ell_a, \nu_a) = M(\ell_b, \nu_b)$, which leads to the constraint $M_{Wa} = M_{Wb}$. Alternatively, the momentum of the dineutrino could also be split according to the minimisation of the following two functions: (iii) $|M_{t_a}^2 + M_{t_b}^2|$, or (iv) $|M_{t_a} - M_{t_b}|$.

Thus the four methods used in this analysis are as follows:

\begin{enumerate}
    \item[$\textbf{A}$:] $M^a_{\rm top} = M^b_{\rm top}$
    \item[$\textbf{B}$:] $M_{W}^a = M_{W}^b$
    \item[$\textbf{C}$:] min $\Sigma M_{\rm top}^2$
    \item[$\textbf{D}$:] min $\Delta M_{\rm top}$
\end{enumerate}

The performance of the four methods can be evaluated on the basis of the $m_{\ttbar}$ variable, which is one of the primary observables  for the phenomenological study. We also compare the $p_{T}^{\rm reco}$ versus $p_{T}^{\rm truth}$. 

\begin{figure*}[htbp]
    \centering
    \includegraphics[width=0.45\linewidth]{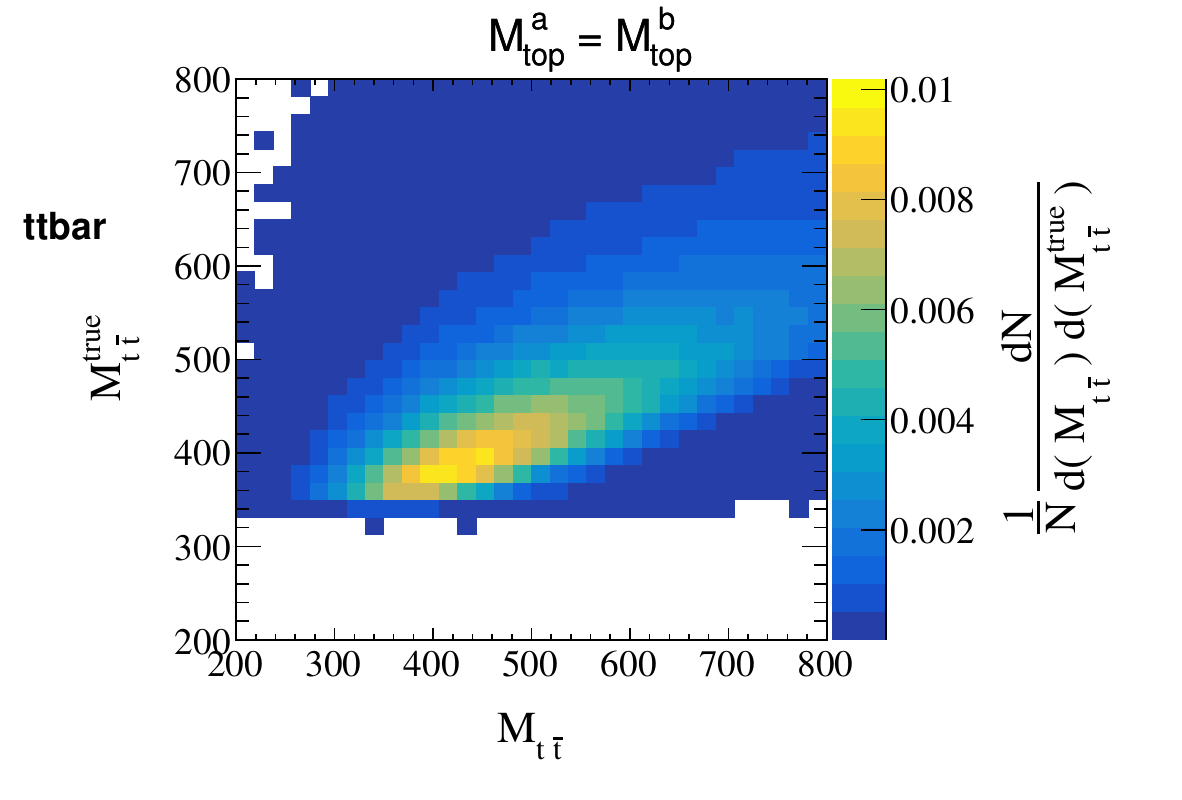}
    \includegraphics[width=0.45\linewidth]{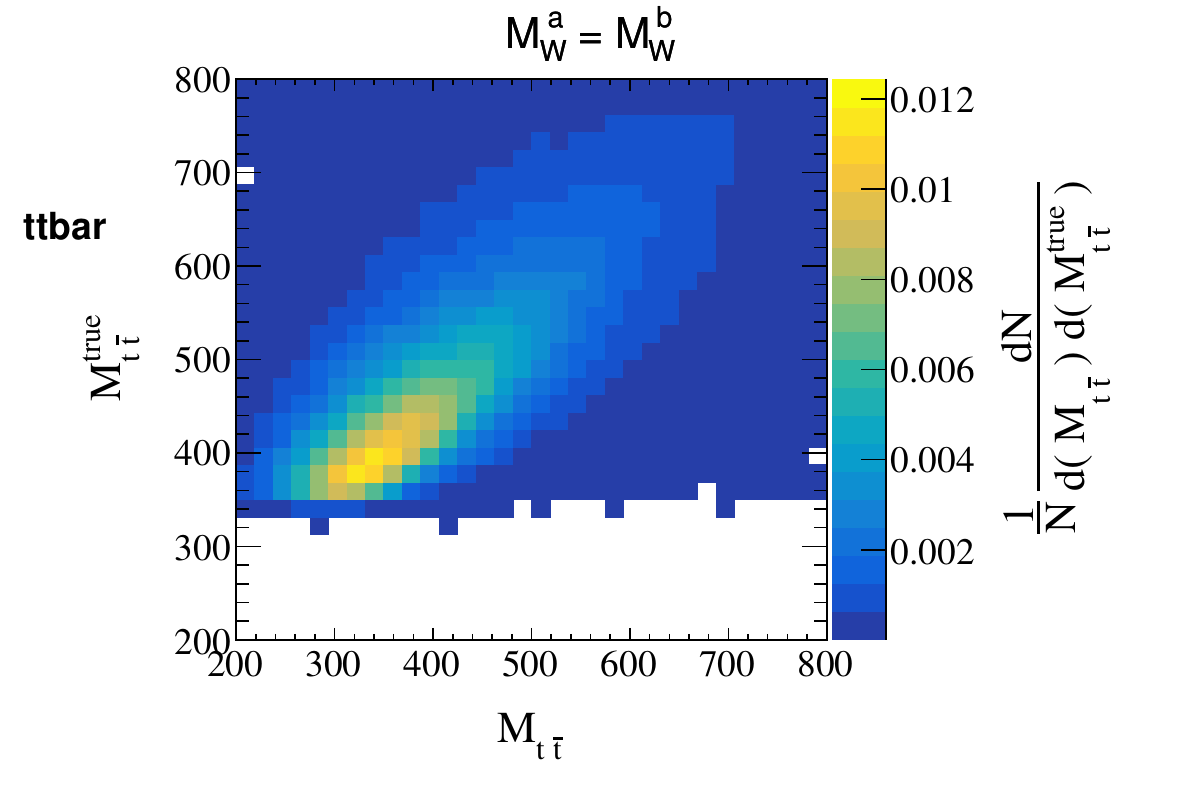}\\
    \includegraphics[width=0.45\linewidth]{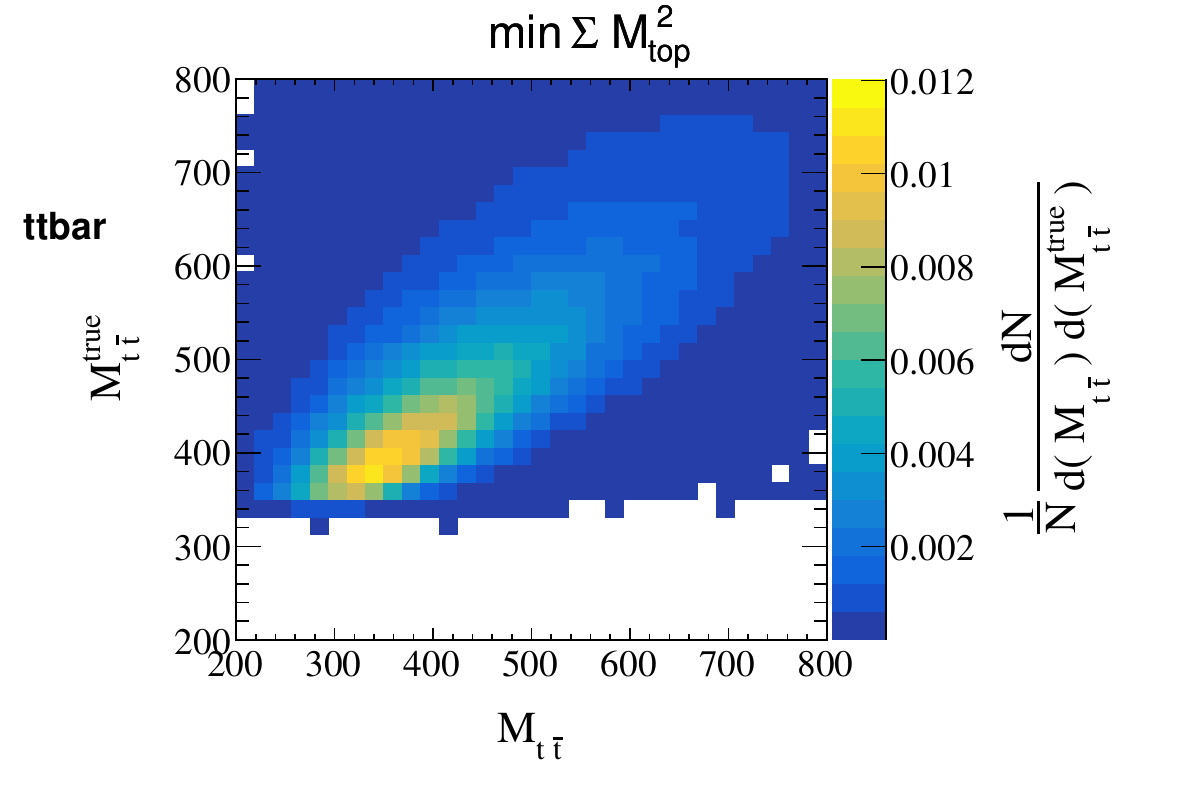}
    \includegraphics[width=0.45\linewidth]{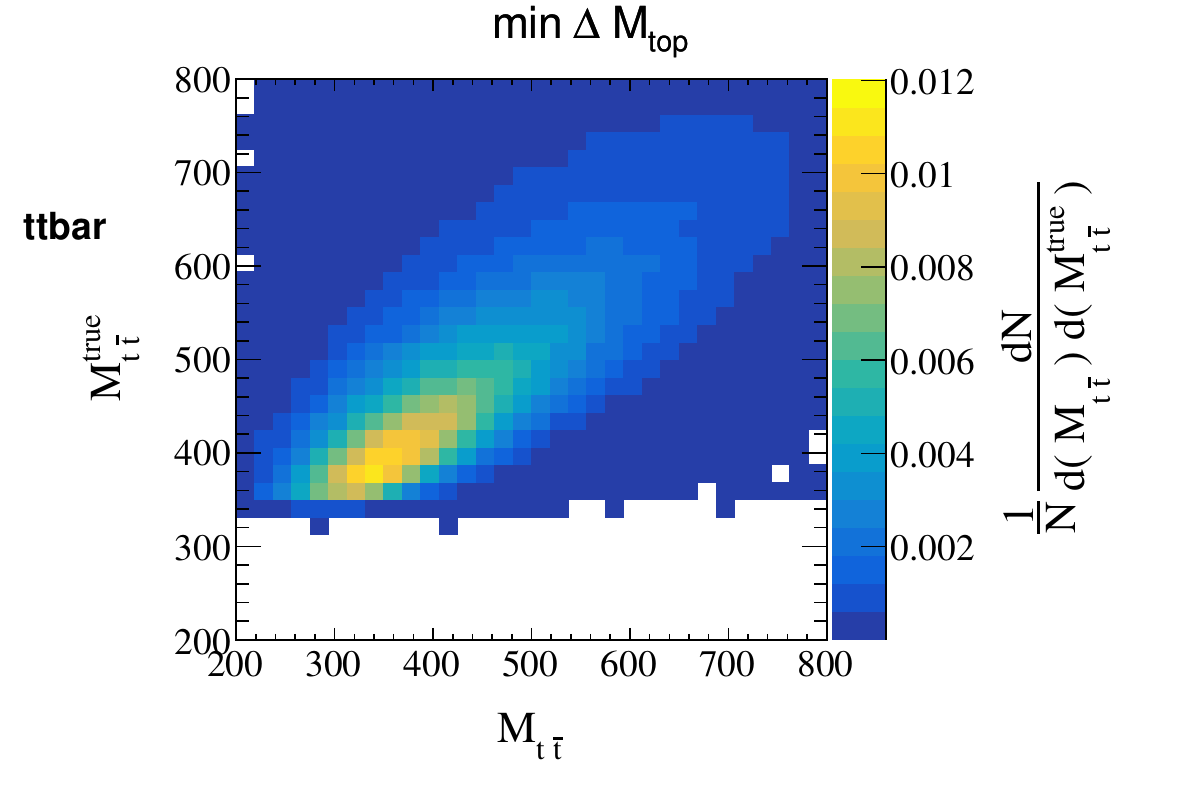}
    \caption{Performance of the reconstruction methods in reconstructing $M_{\ttbar}$ (a) Method A (b) Method B (c) Method C and (d) Method D after pre-selection criteria are applied.}
    \label{fig:2dcompare_mtt}
\end{figure*}

\begin{figure*}[htbp]
    \centering
    \includegraphics[width=0.45\linewidth]{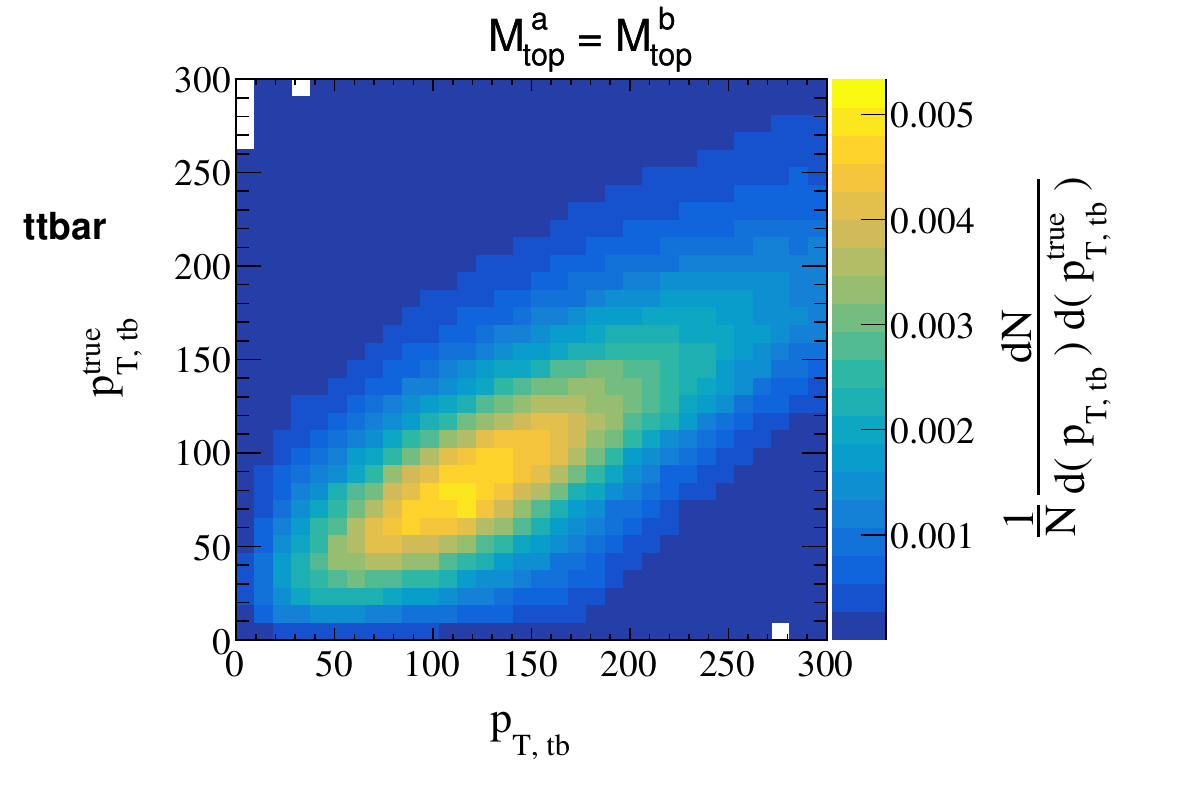}
    \includegraphics[width=0.45\linewidth]{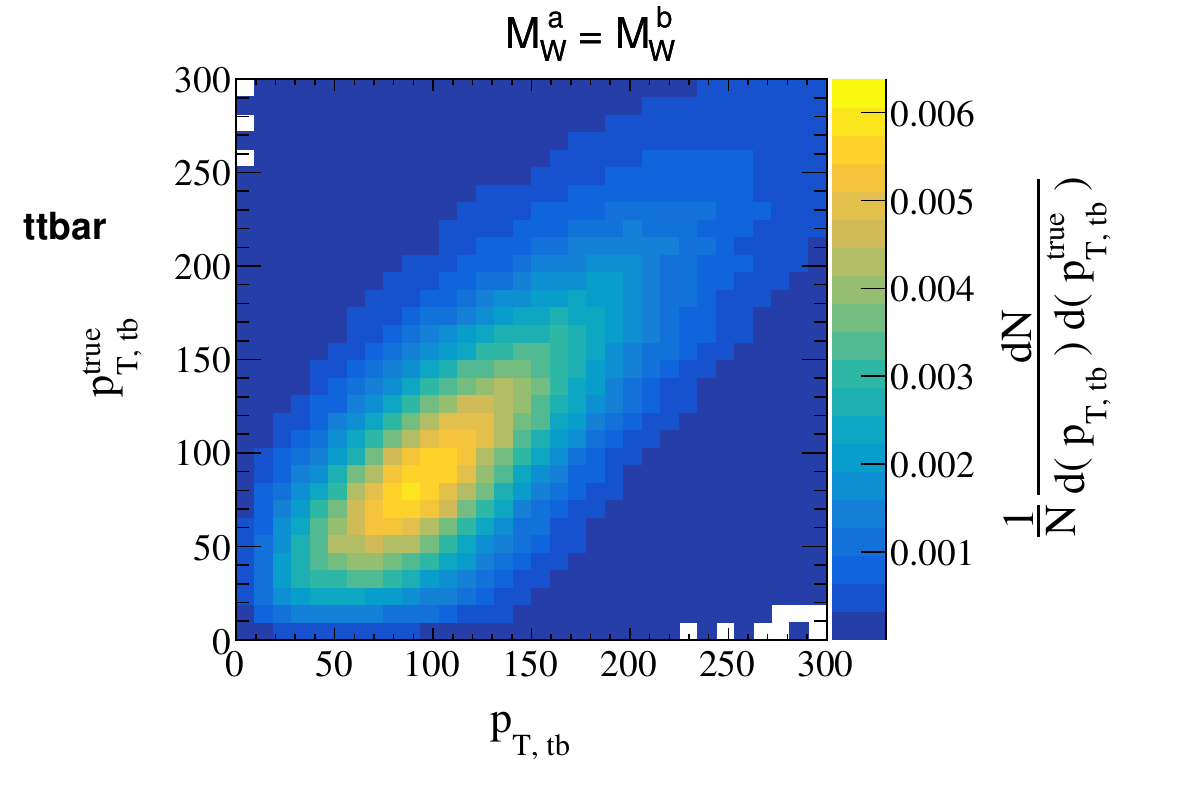}\\
    \includegraphics[width=0.45\linewidth]{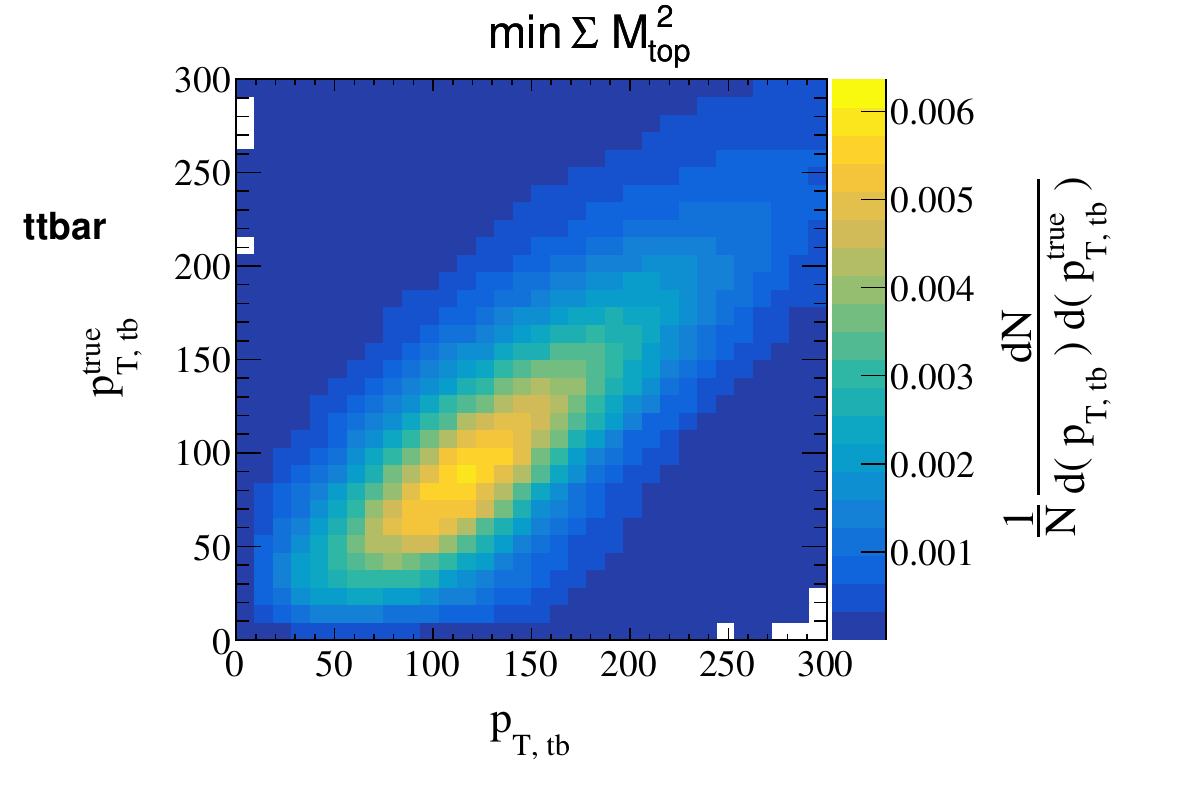}
    \includegraphics[width=0.45\linewidth]{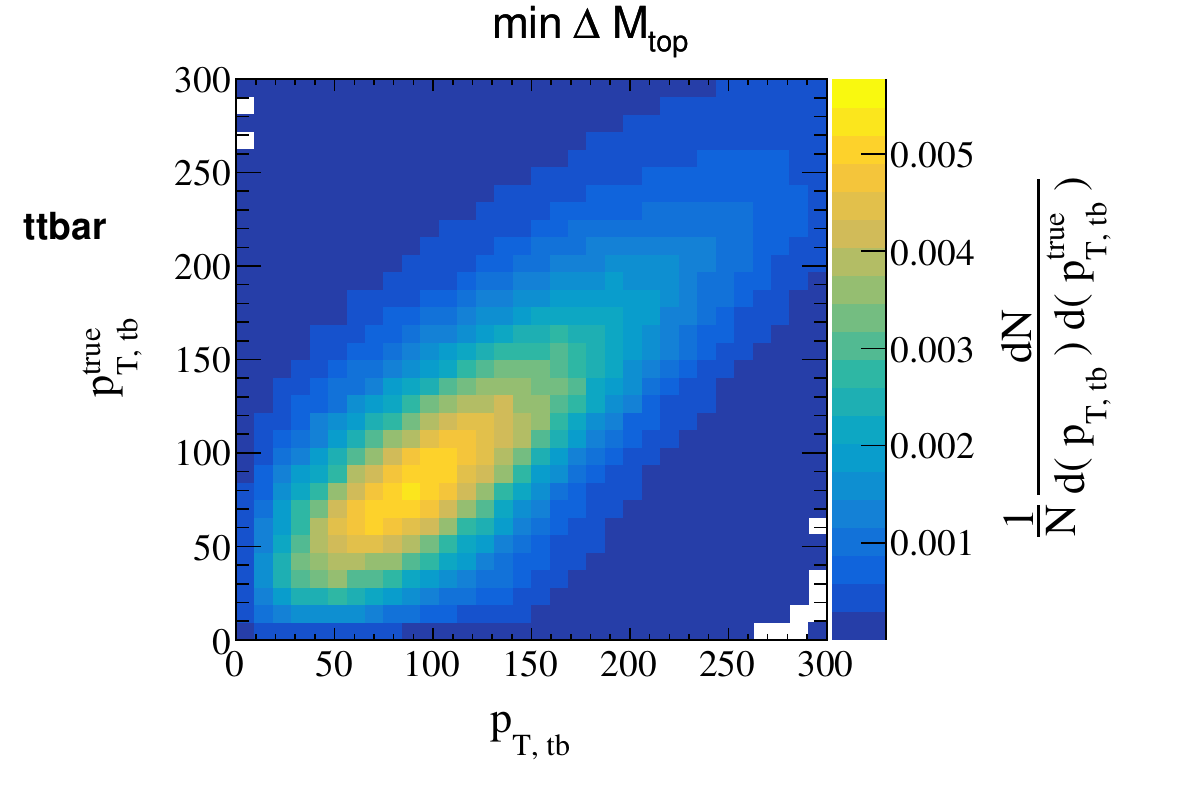}
    \caption{Performance of the reconstruction methods in reconstructing $p_T$ of top quark (a) Method A (b) Method B (c) Method C and (d) Method D after pre-selection criteria are applied.}
    \label{fig:2dcompare_pt}
\end{figure*}

\begin{figure*}[htbp]
    \centering
    \includegraphics[width=0.45\linewidth]{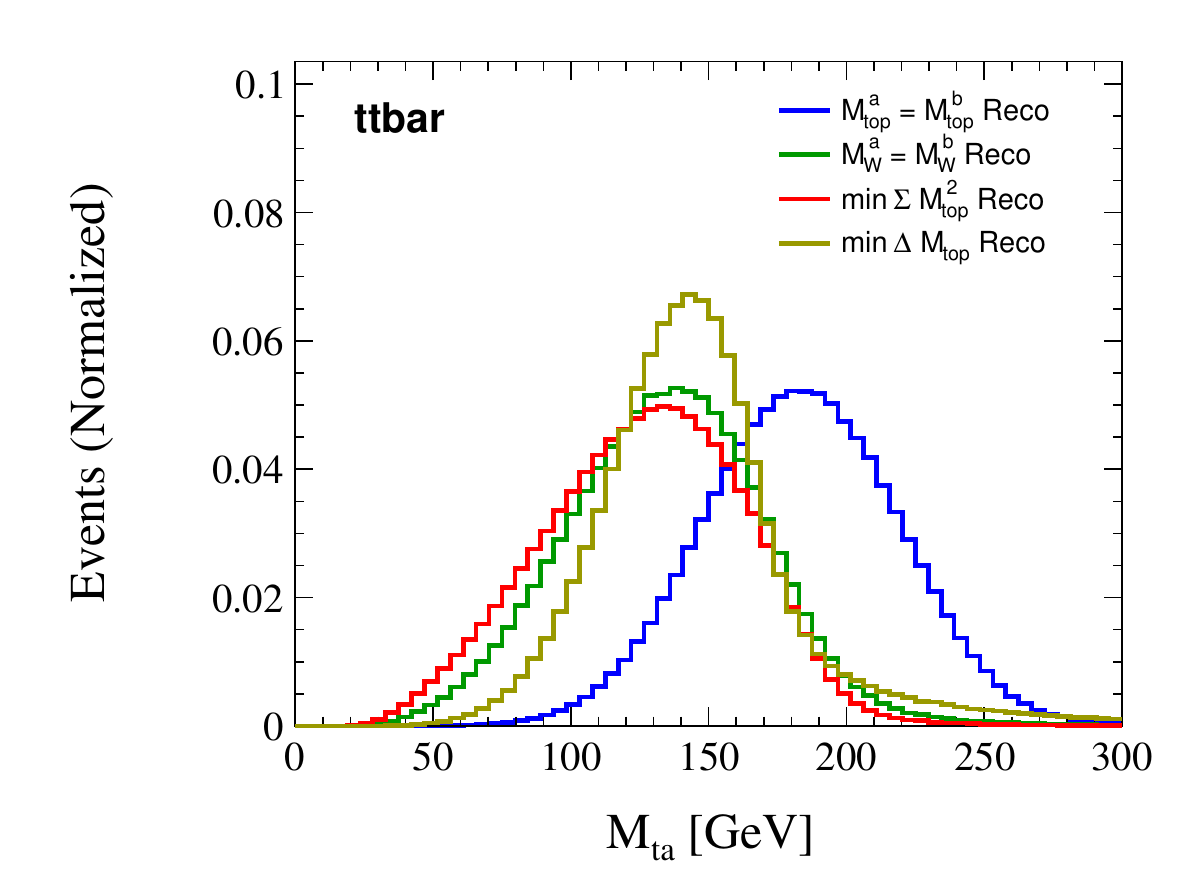}
    \includegraphics[width=0.45\linewidth]{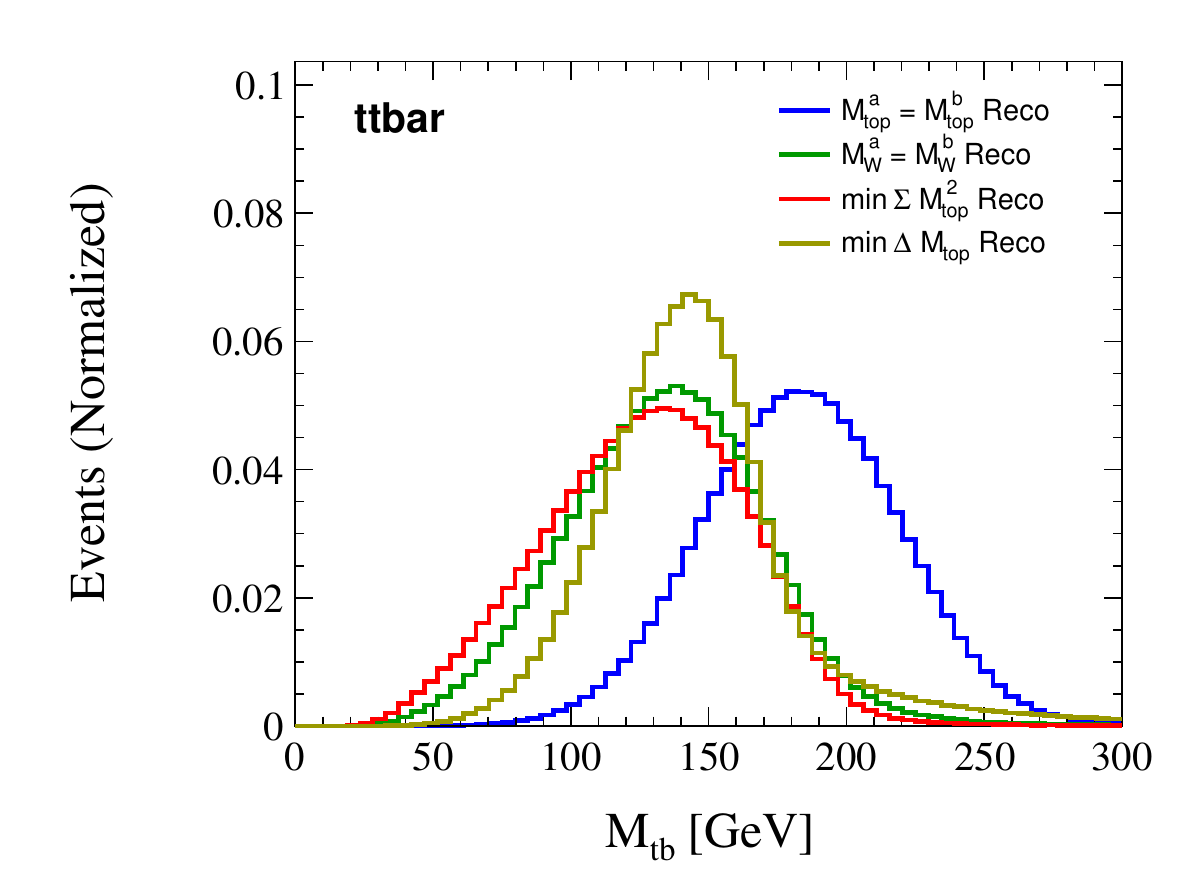}\\
    \caption{Top quark mass as reconstructed by the four methods used in this analysis. Mass of top quark labelled as (Left) $ta$ and (Right) $tb$ in the \RF~package after pre-selection criteria are applied.}
    \label{fig:topmass}
\end{figure*}

\begin{figure*}[htbp]
    \centering
    \includegraphics[width=0.45\linewidth]{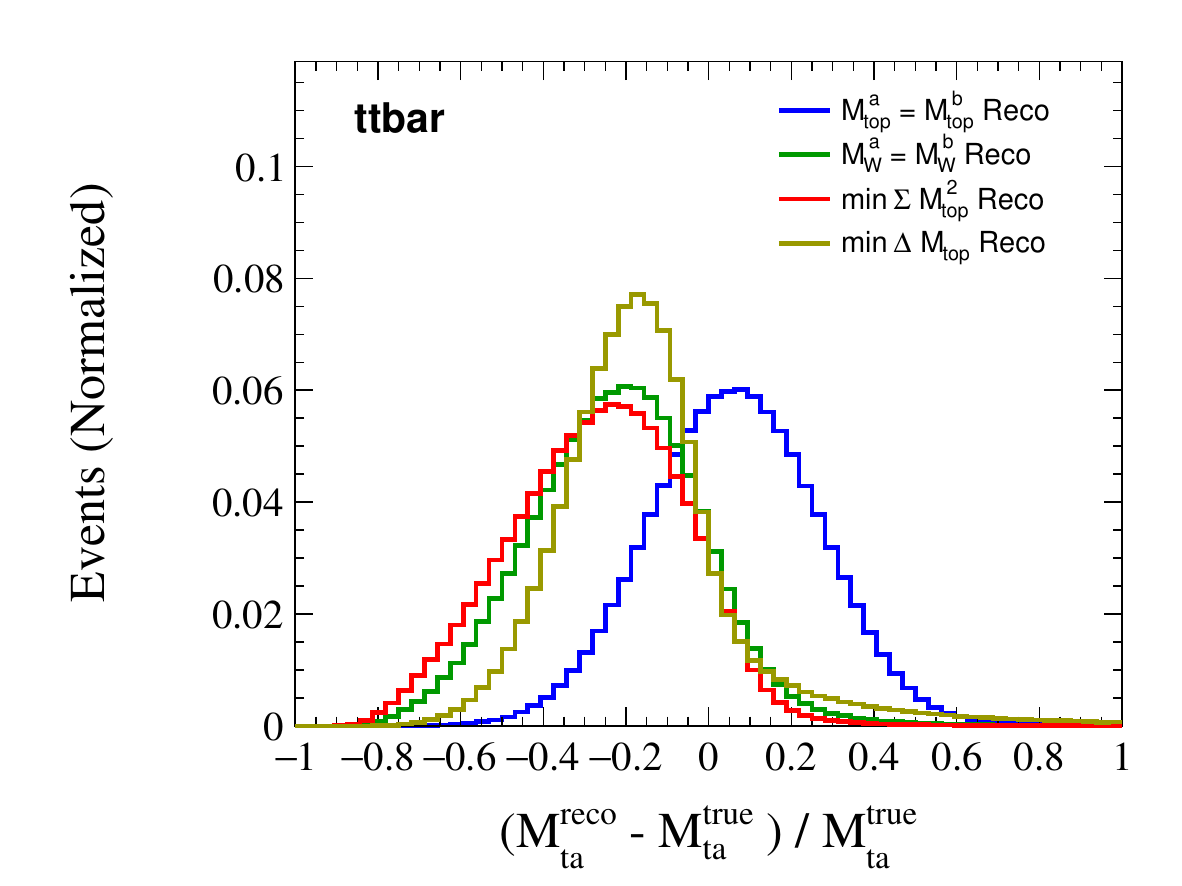}
    \includegraphics[width=0.45\linewidth]{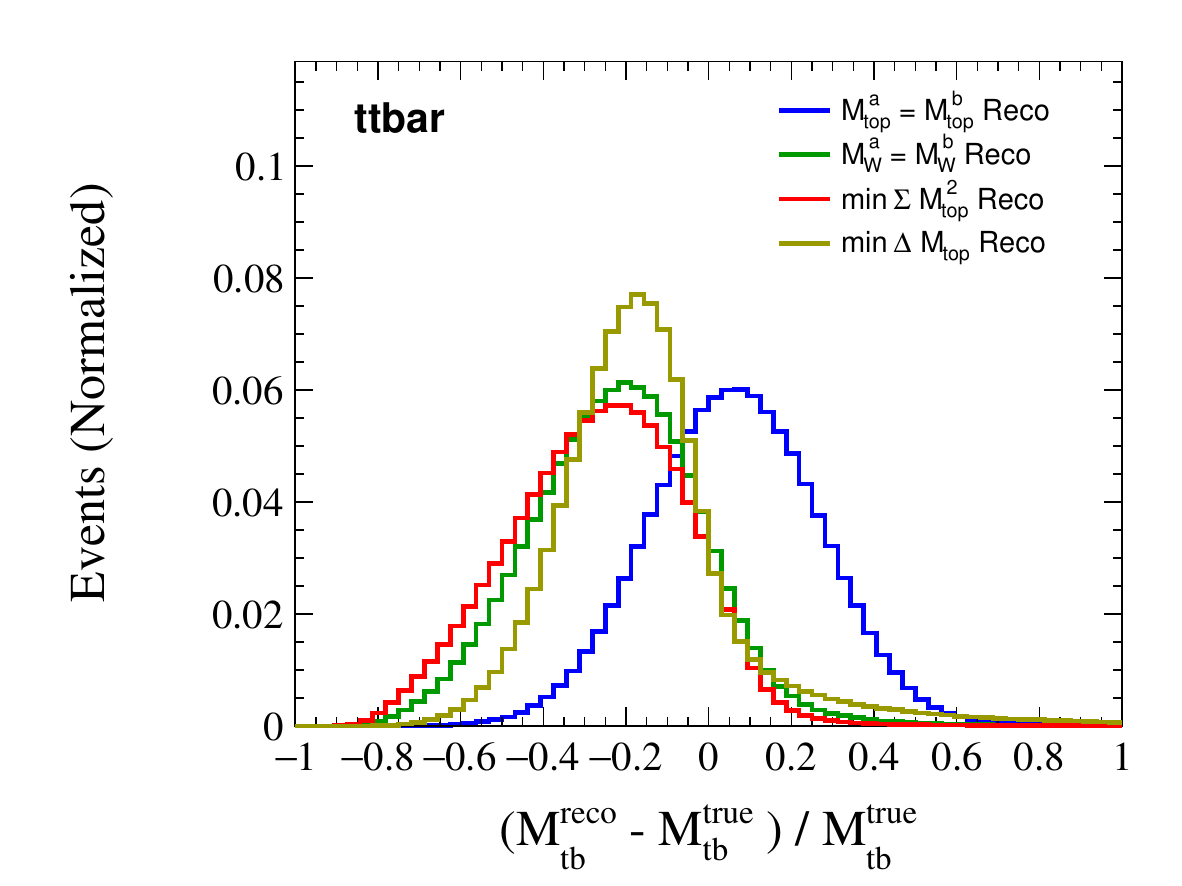}\\
    \includegraphics[width=0.45\linewidth]{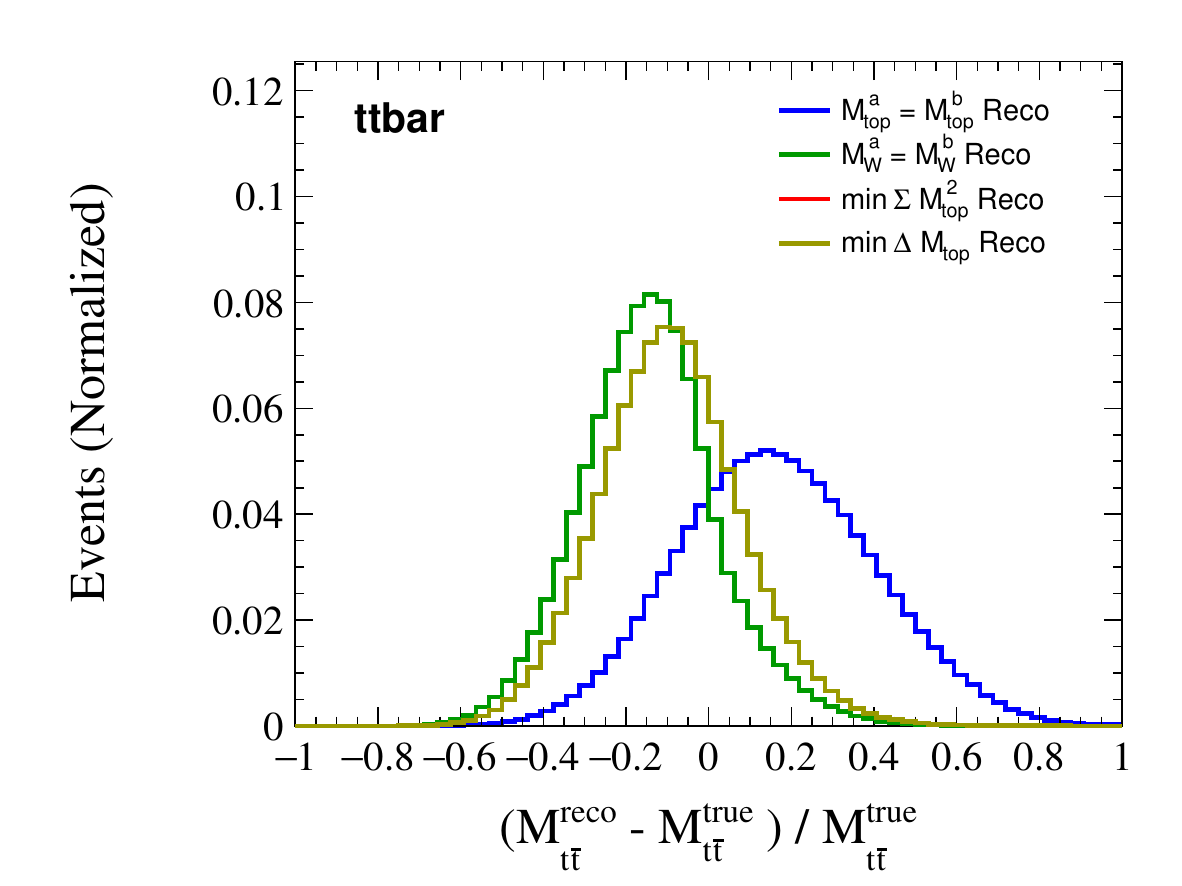}
    \caption{Resolution of (a) $M_{ta}$ (b) $M_{tb}$  and (c) $M_{t\bar{t}}$ for the four reconstruction methods in after pre-selection criteria are applied.}
    \label{fig:resolution}
\end{figure*}

All four methods indicate a correlation for the $M_{\ttbar}$ variable in Figure~\ref{fig:2dcompare_mtt} when compared against the truth value. The same observation is made from the top quark's transverse momentum distribution in Figure~\ref{fig:2dcompare_pt}. The distribution of the top quark mass for different reconstruction methods is given in Figure~\ref{fig:topmass}. We also compare the resolution for each of the four methods for the variables such as the top quark masses ($t_a$ and $t_b$), and $M_{t\bar{t}}$ variable in \autoref{fig:resolution}. From these results, we see that the optimal method to allow for a consistent top mass would be Reconstruction method A.

\begin{figure*}[htbp]
    \centering
    \includegraphics[width=0.45\linewidth]{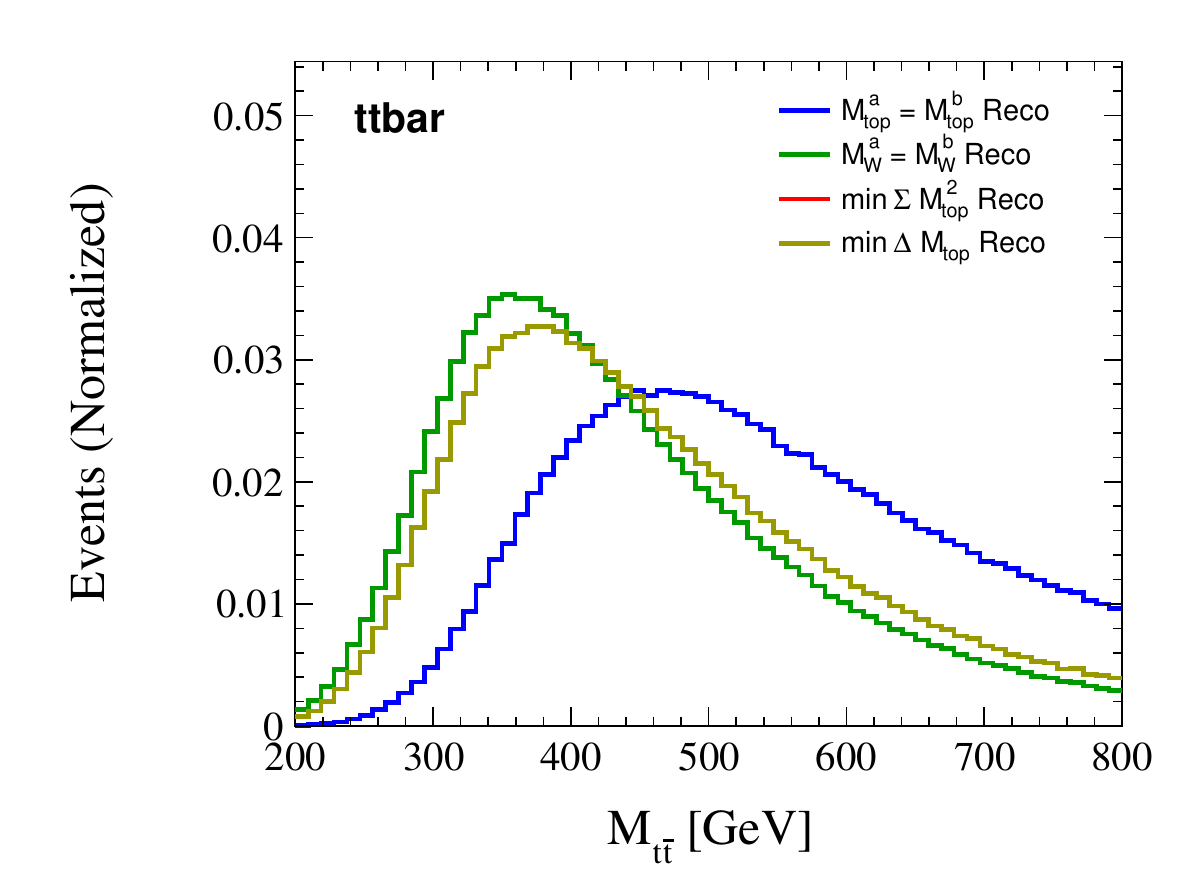}
    \includegraphics[width=0.45\linewidth]{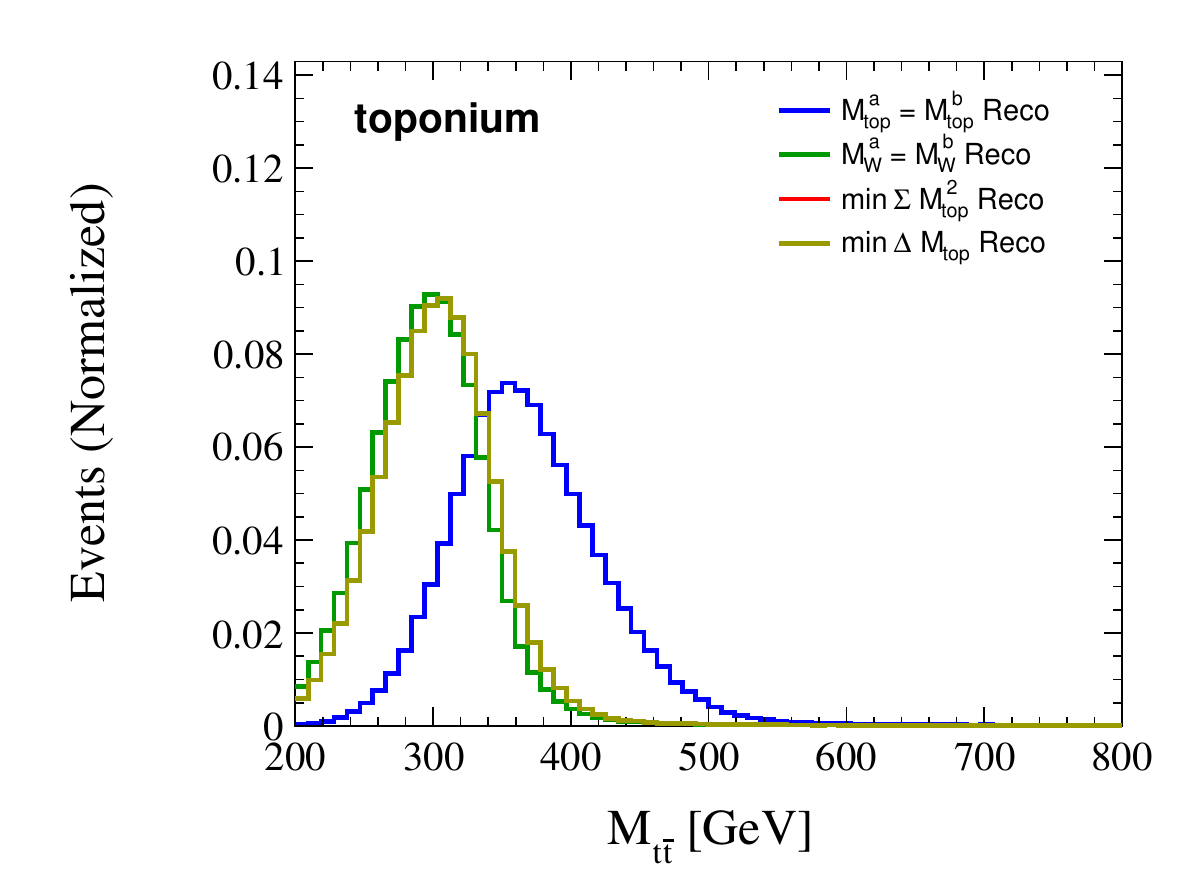}
    \caption{Invariant mass of the $\ttbar$ as computed by the reconstruction algorithms after pre-selection criteria are applied.}
    \label{fig:mtt}
\end{figure*}

The $\ttbar$~invariant mass distribution as reconstructed using the \RF~software~\cite{Jackson:2017gcy} is given in Figure~\ref{fig:mtt}. As expected, the algorithm is able to capture the peak structure of the $\ttbar$ invariant mass in the toponium sample. 

\section{Analysis using Pre-selection Criteria}\label{sec:preanalysis}

\begin{figure*}
    \centering
    \includegraphics[width=0.45\linewidth]{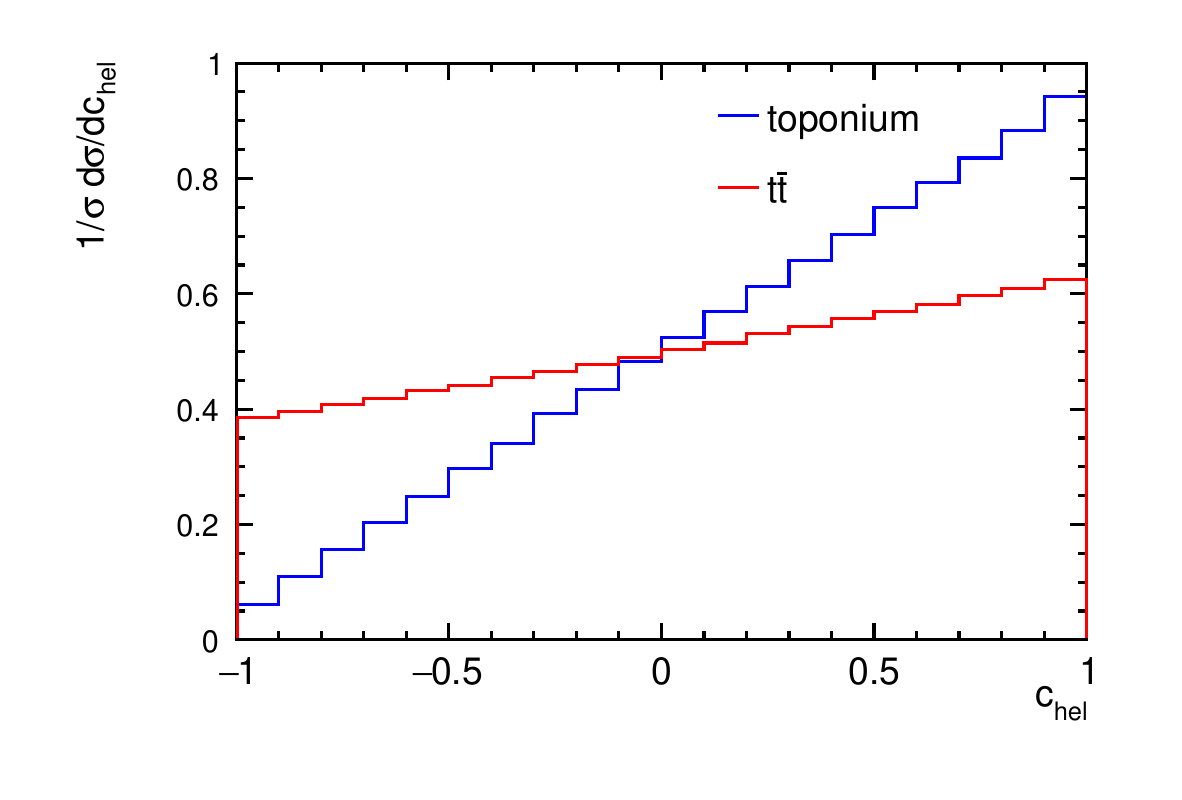}
    \includegraphics[width=0.45\linewidth]{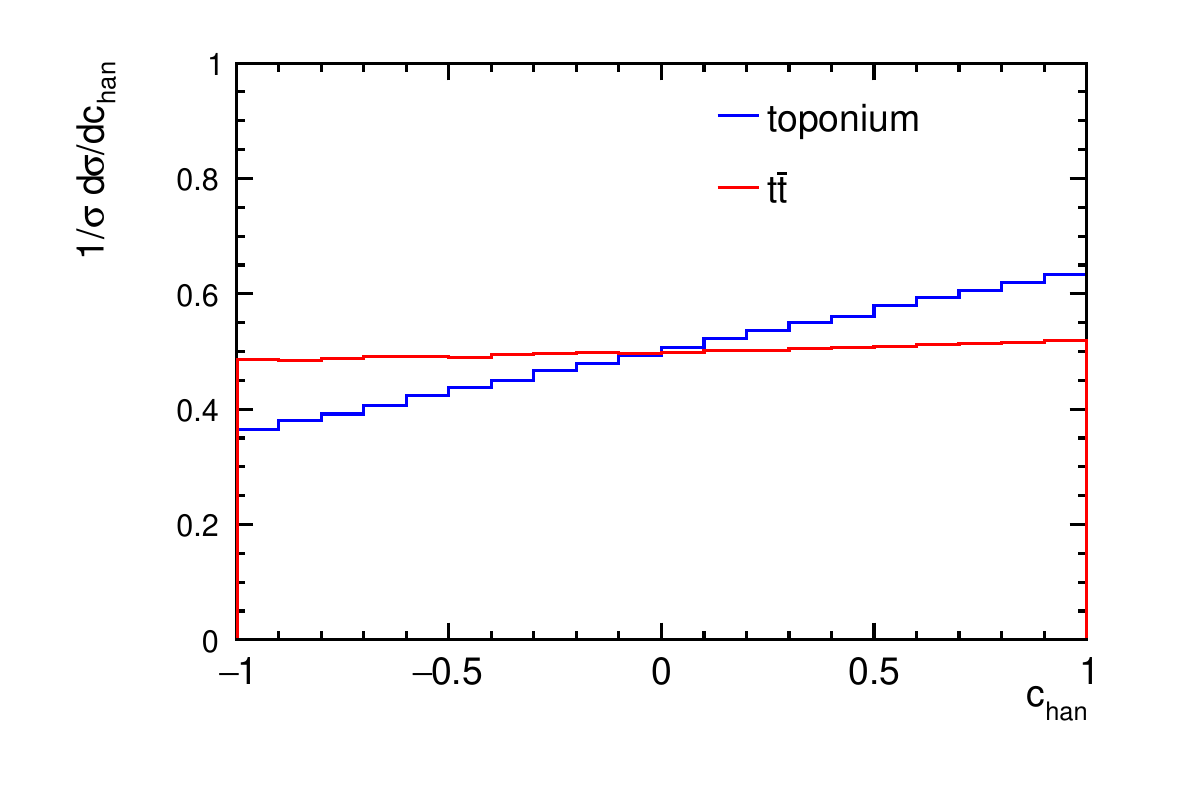}
    \caption{The distribution of $\chel$ and $\chan$ variables before applying pre-selection criteria at truth level.}
    \label{fig:chanchel_nocut}
\end{figure*}

\begin{figure*}
    \centering
    \includegraphics[width=0.45\linewidth]{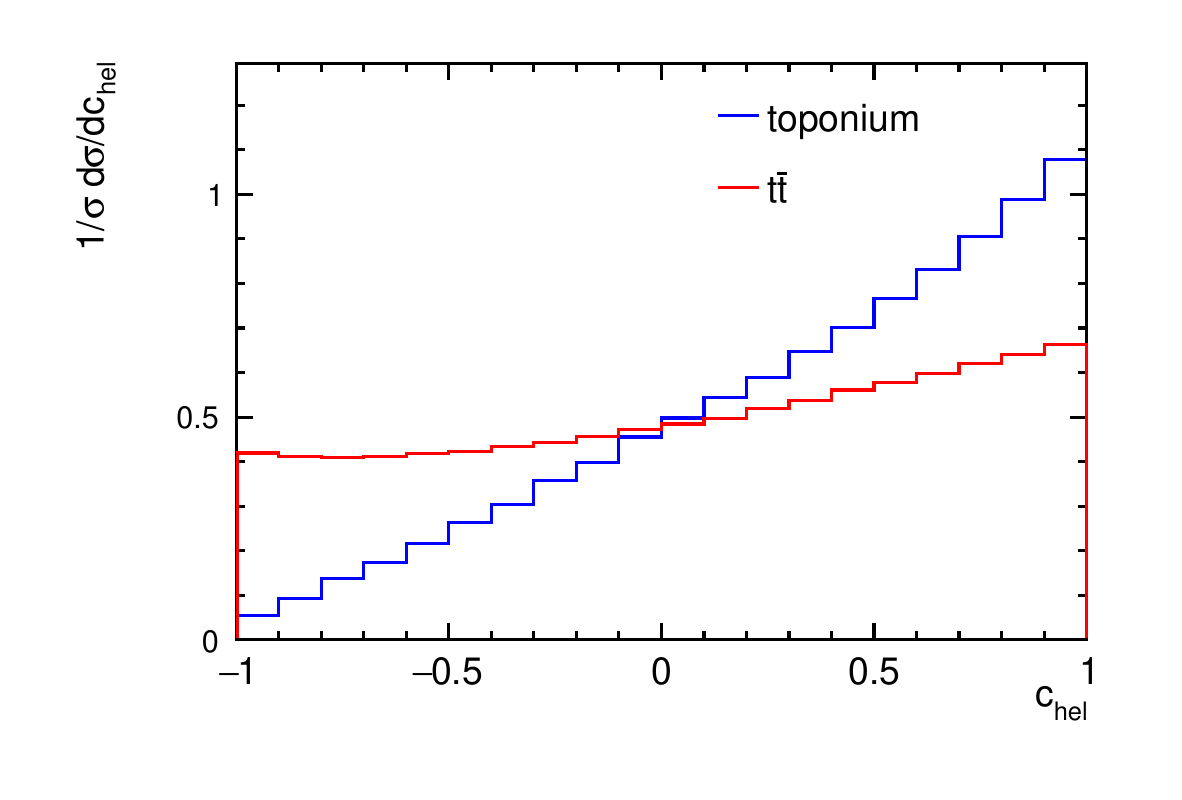}
    \includegraphics[width=0.45\linewidth]{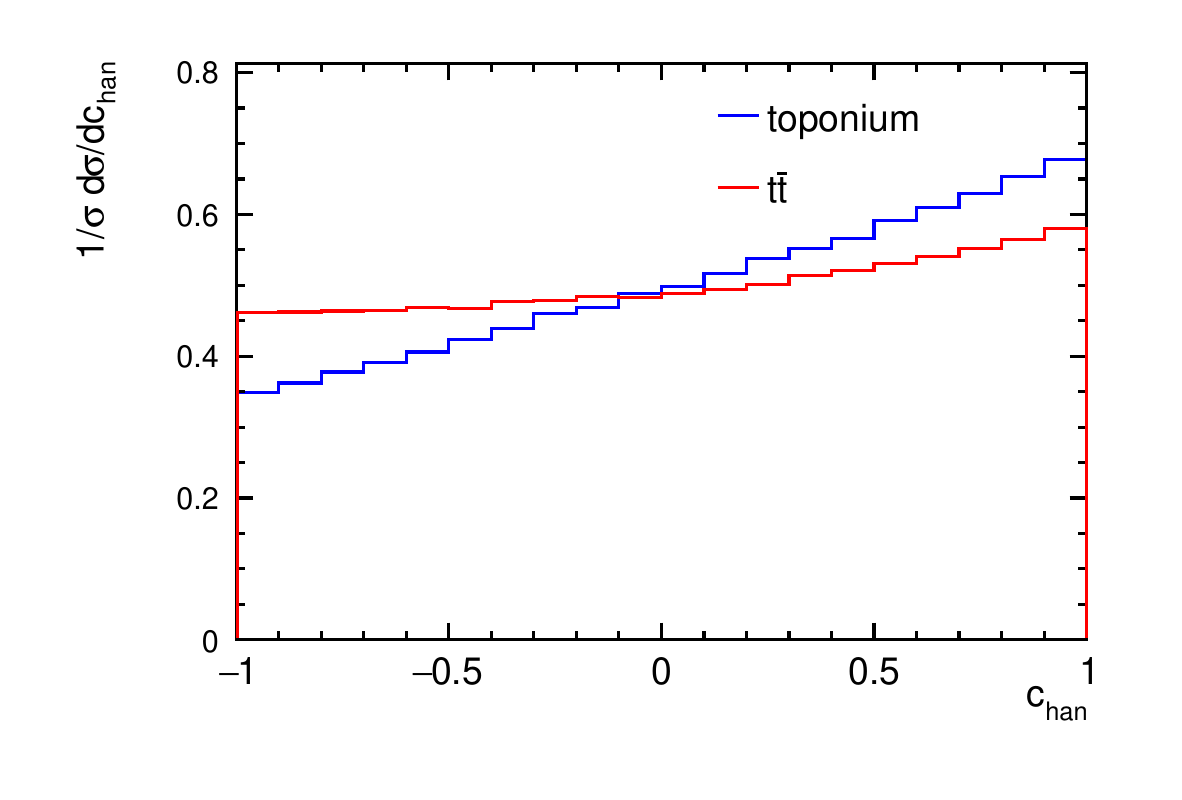}
    \caption{The distribution of $\chel$ and $\chan$ variables after applying pre-selection criteria at truth level.}
    \label{fig:chanchel}
\end{figure*}

In the ATLAS and CMS analyses, nine regions formed by $\chel$ and $\chan$ variables~\cite{ATLAS:2026dbe,CMS:2025kzt} were used to maximise sensitivity for the toponium signal. The $\chel$ variable is constructed by taking the dot product of the lepton's momenta, which is first boosted to the $\ttbar$ CoM frame of reference. The individual leptons are boosted to the parent top and antitop quark rest frames, respectively, where these top quarks have already been boosted to the $\ttbar$ CoM frame. The $\chan$~variable is constructed in the same manner but flipping the sign of one of the lepton's momentum along the direction of the top quark. These variables were originally used in other analyses by the collaborations~\cite{Aguilar-Saavedra:2022uye,ATLAS:2023fsd,CMS:2024pts,CMS:2019nrx,CMS:2019pzc,Bernreuther:2004jv}. The nine regions are formed by considering $-1 < \chel < -\frac{1}{3}$,  $-\frac{1}{3} < \chel < \frac{1}{3}$, $\frac{1}{3} < \chel < 1$ and the same for $\chan$. At the truth level, $\chel$ and $\chan$ distributions for the present analysis are shown in Figure~\ref{fig:chanchel_nocut} before applying any preselections and in Figure~\ref{fig:chanchel} after applying the preselection criteria. As seen in \autoref{fig:chanchel_nocut}, before applying any preselections, $\chel$ variable shows a maximal slope for the toponium sample, whereas for the $\chan$ variable, the distribution exhibits nearly flat shape for the baseline $\ttbar$~sample, whereas for the toponium sample, the slope is steeper.

Mathematically the variables are defined as~\cite{Aguilar-Saavedra:2022uye}: 

\begin{align}
& \hat{p}_{a} = (\sin \theta_a \cos \varphi_a,\sin \theta_a \sin \varphi_a,\cos \theta_a ) \,, \notag \\
& \hat{p}_{b} = (\sin \theta_b \cos \varphi_b,\sin \theta_b \sin \varphi_b,\cos \theta_b ) \,.
\label{papb}
\end{align}

where $\theta_{a,b}$ and $\varphi_{a,b}$ are the polar and azimuthal angles, respectively, of the unit momentum vectors $\hat{p}_{a,b}$. The $\hat{}$ represents that the variables are boosted according to the definition above. The scalar product $\hat{p}_{a} \cdot \hat{p}_{b}$ gives the $\chel$ variable. In order to obtain $\chan$, we flip the sign of momentum for one of the lepton for the component parallel to the top quark direction and evaluate the scalar dot product as before.

\begin{figure}[!htbp]
    \centering
    \includegraphics[width=0.90\linewidth]{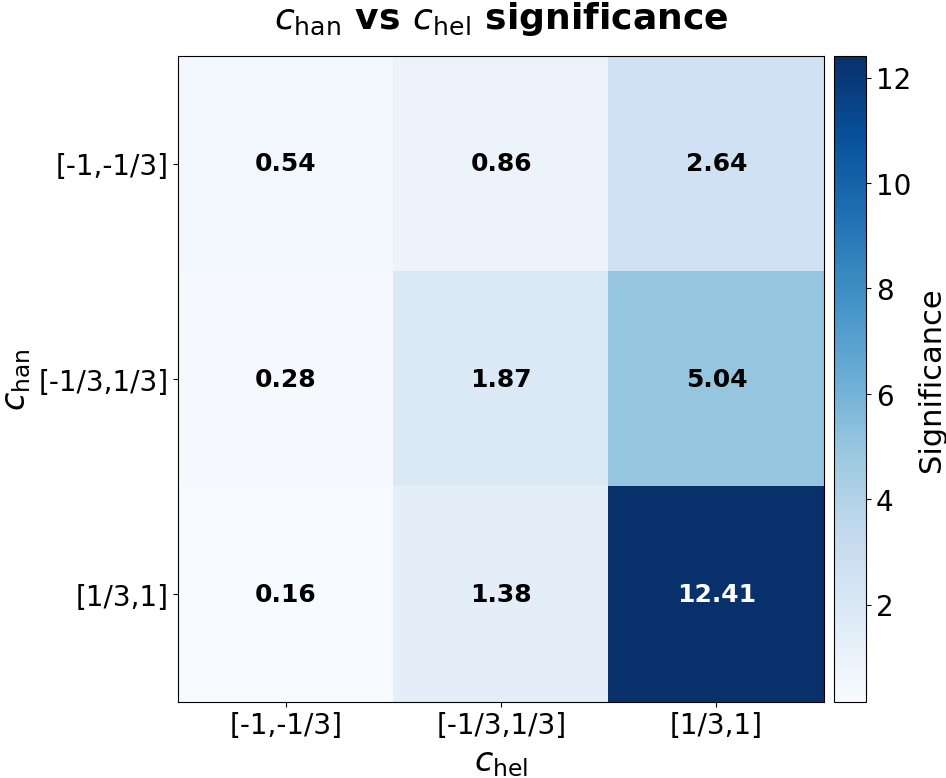}
    \caption{The significance achieved in the nine analysis regions. The optimal region has a significance of $12.4\sigma$ following pre-selection criteria.}
    \label{fig:chanchel_sig}
\end{figure}

\begin{figure}[!htbp]
    \centering
    \includegraphics[width=0.90\linewidth]{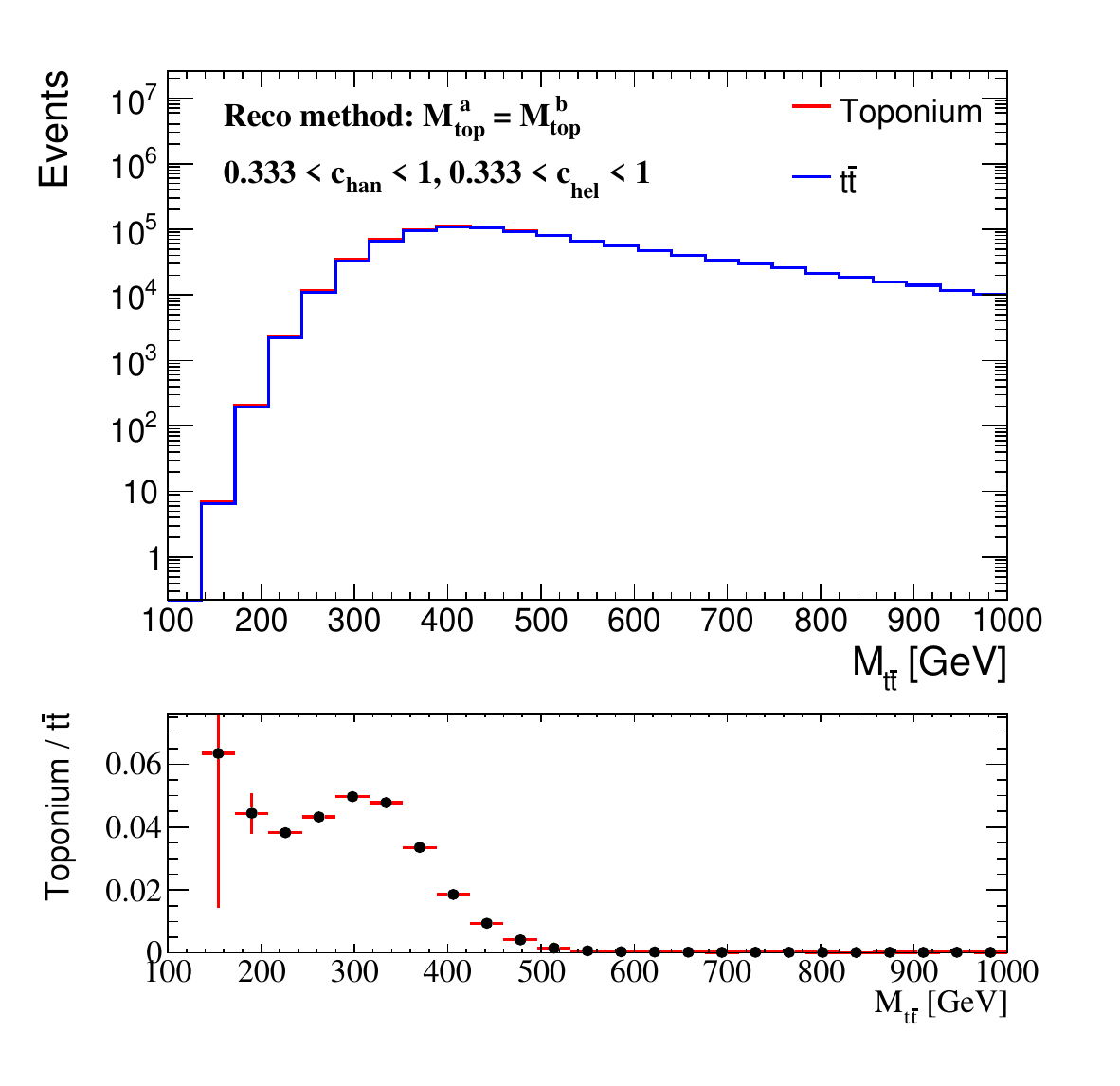}
    \caption{The invariant mass distribution of $\ttbar$ in the optimal region following pre-selection criteria. }
    \label{fig:chanchel_dist}
\end{figure}

We consider the Reconstruction method A. The analysis is carried out over the defined $\chel$ and $\chan$ bins and our results are presented in the form of significance achieved in each region Figure~\ref{fig:chanchel_sig}. The stacked distributions in the region, which is $\chel\in (1/3,1)$ and $\chan\in (1/3,1)$ (optimal region) is given in Figure~\ref{fig:chanchel_dist}.

The significance metric is defined as $\frac{S}{\sqrt{S+B}}$ where $S$ represents the yield from the toponium sample and $B$ represents the baseline yield of the $\ttbar$ sample which is used to assess the sensitivity of the analysis. We note that the significance achieved with this method at the LHC Run 3 energy, luminosity and the binning strategy as above, in the optimal region is $\approx 12.4\sigma$. Using the same analysis procedure, the significance expected to be achieved using LHC Run 2 energy and luminosity is found to be $8.4\sigma$.

\begin{figure}[htbp]
    \centering
    \includegraphics[width=0.90\linewidth]{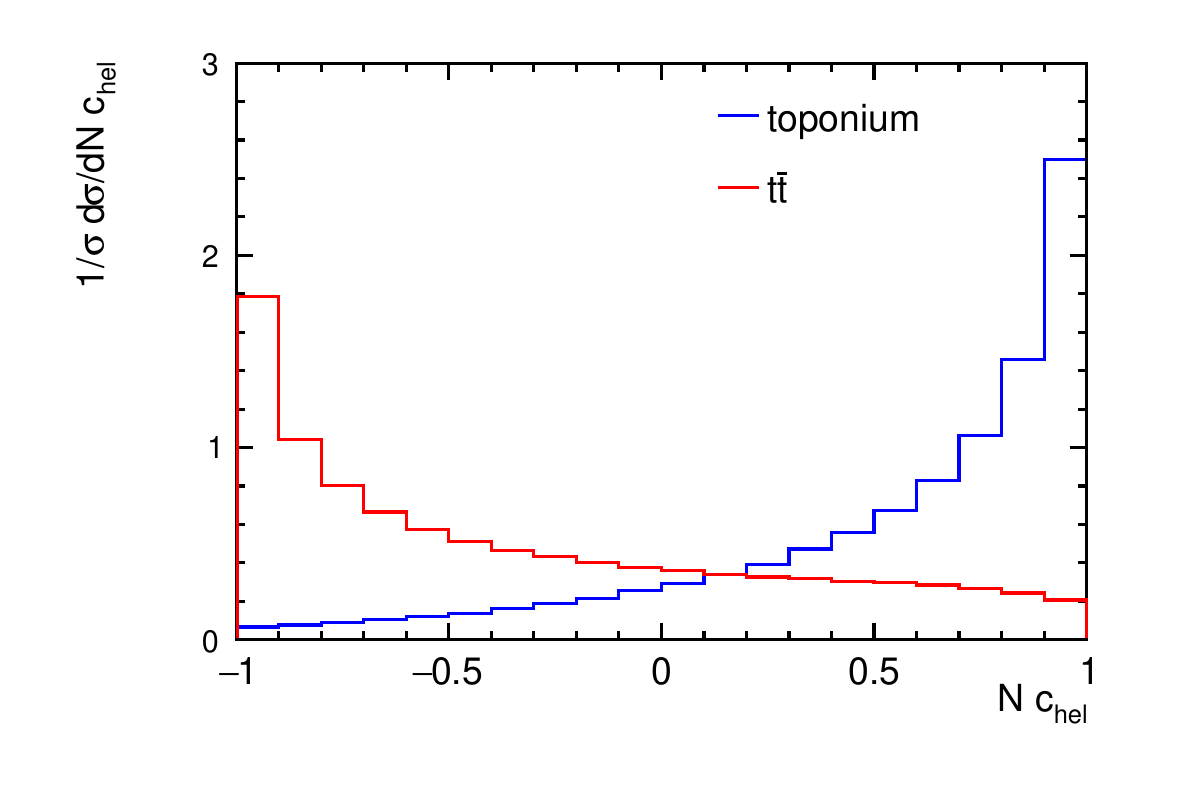}\\
    \caption{The distribution of  the $\nchel$ variable after pre-selection criteria at the truth level.}
    \label{fig:nchel}
\end{figure}

In this study we propose two variables which could be used in improving the sensitivity in the toponium analysis for the dilepton final state. The variables presented here are named: $\Delta\phi(t,\bar{t})$ which is the difference in azimuthal angle of a reconstructed top quark pair and $\nchel$ which is defined in a similar fashion as the $\chel$ variable using a different frame of reference. In particular, the top quark or anti-top quark is not boosted to the $\ttbar$ frame of reference when applying the boosts involved in computing $\chel$. The parton-level distributions are presented in Figure~\ref{fig:nchel}. The correlations between the $\nchel$ variable with the $\Delta\phi(t,\bar{t})$ variable for the $\ttbar$ and toponium samples are also presented.

\begin{figure*}[!htbp]
    \centering
    \includegraphics[width=0.48\linewidth]{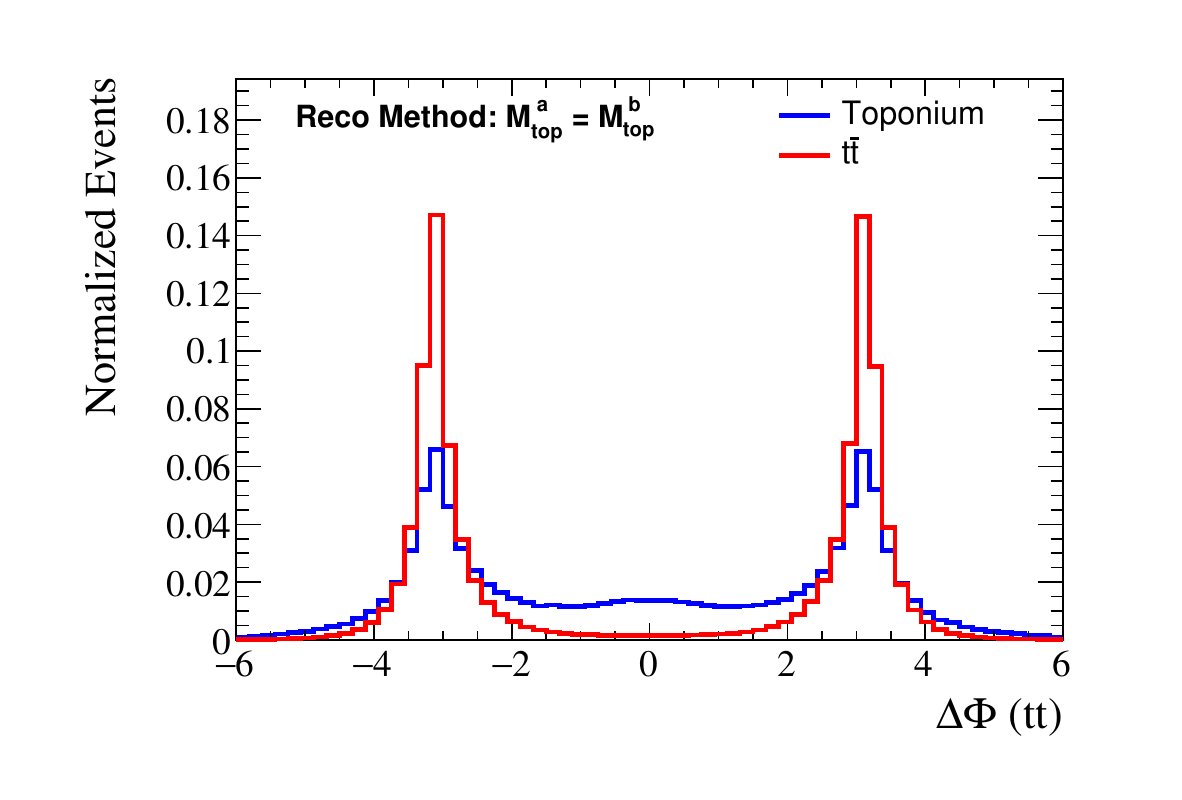}
    \includegraphics[width=0.48\linewidth]{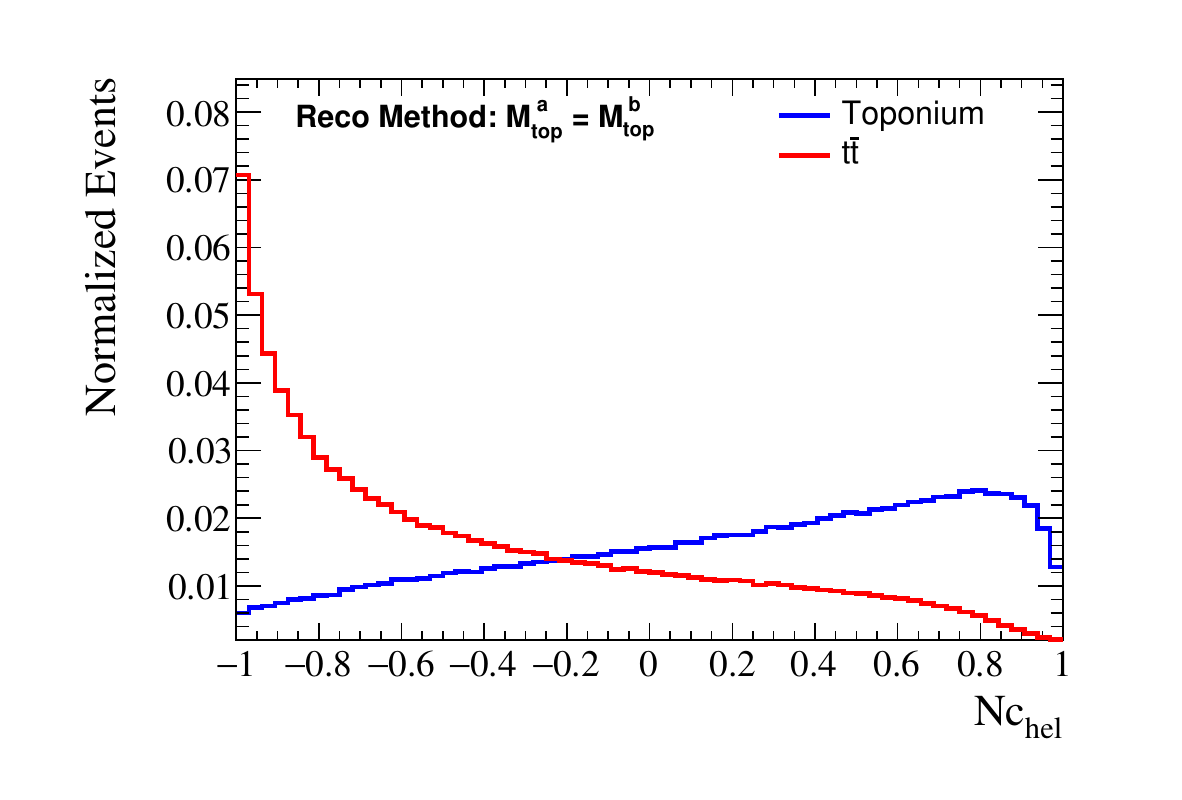}
    \caption{The distribution of the $\Delta\Phi(\ttbar)$ and the $\nchel$ variables after pre-selection criteria at the reconstructed level.}
    \label{fig:nchel_reco1}
\end{figure*}

The reconstruction of both these proposed variables rely on the ability to reconstruct the top quarks. In order to achieve that we apply, as before, method A for reconstruction and find the following distribution at the reconstruction level (Figure~\ref{fig:nchel_reco1}). The distributions are consistent with the observation concluded at the parton level. 

As a next step, we take the Cartesian product of three regions in each variable, forming nine phase-space regions. Note that the regions are not optimised and that the criteria applied are motivated by visual inspection of the distributions. We consider the following bins: 

\begin{align}
\Delta\Phi(\ttbar) &\in \{[-6,-2],\, [-2,2],\, [2,6]\}, \\
\nchel &\in \{[-1,-0.4],\, [-0.4,0.4],\, [0.4,1]\}, \\[6pt]
\mathcal{R} &= \Delta\Phi(\ttbar) \text{bins} \otimes \nchel\text{bins}.
\end{align}

\begin{figure}[!htbp]
    \centering
    \includegraphics[width=0.9\linewidth]{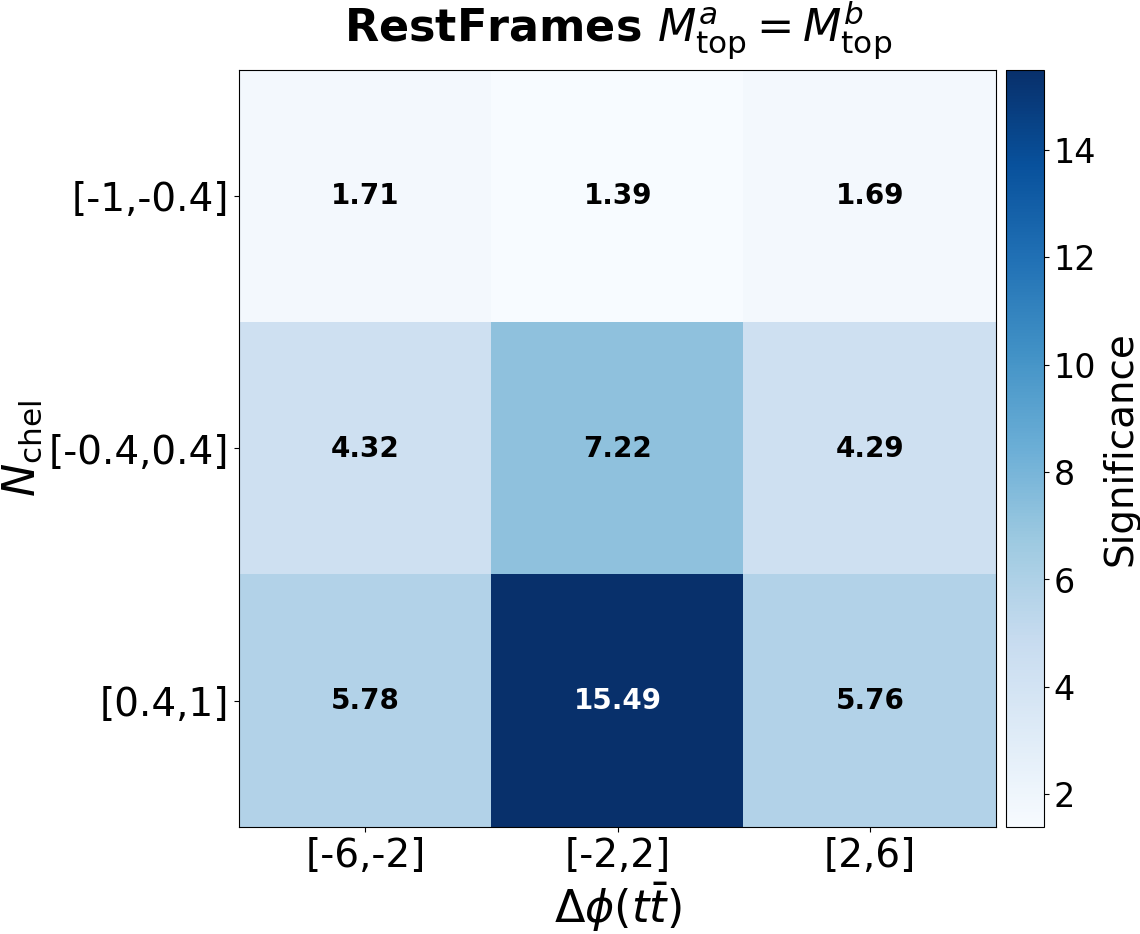}
    \caption{The significance achieved in the nine analysis regions. The optimal region has a significance of $15.5\sigma$ following pre-selection criteria.}
    \label{fig:nchel_sig}
\end{figure}

The significance for observing toponium, using $M_{\ttbar}$ as the observable of interest, is evaluated. The significance is evaluated according to $\frac{S}{\sqrt{S+B}}$ where $S$ represents toponium events and $B$ represents the baseline $\ttbar$ events. The study leads to the following result (Figure~\ref{fig:nchel_sig}). Unlike the $\chel \otimes \chan$ bins where the significance exceeds $5\sigma$ for two bins, in this case the significance is greater than $5\sigma$ in four bins and it is greater than $3\sigma$ in six bins.

\begin{figure*}[htbp]
    \centering
    \begin{subfigure}{0.32\textwidth}
        \centering
        \includegraphics[width=\linewidth]{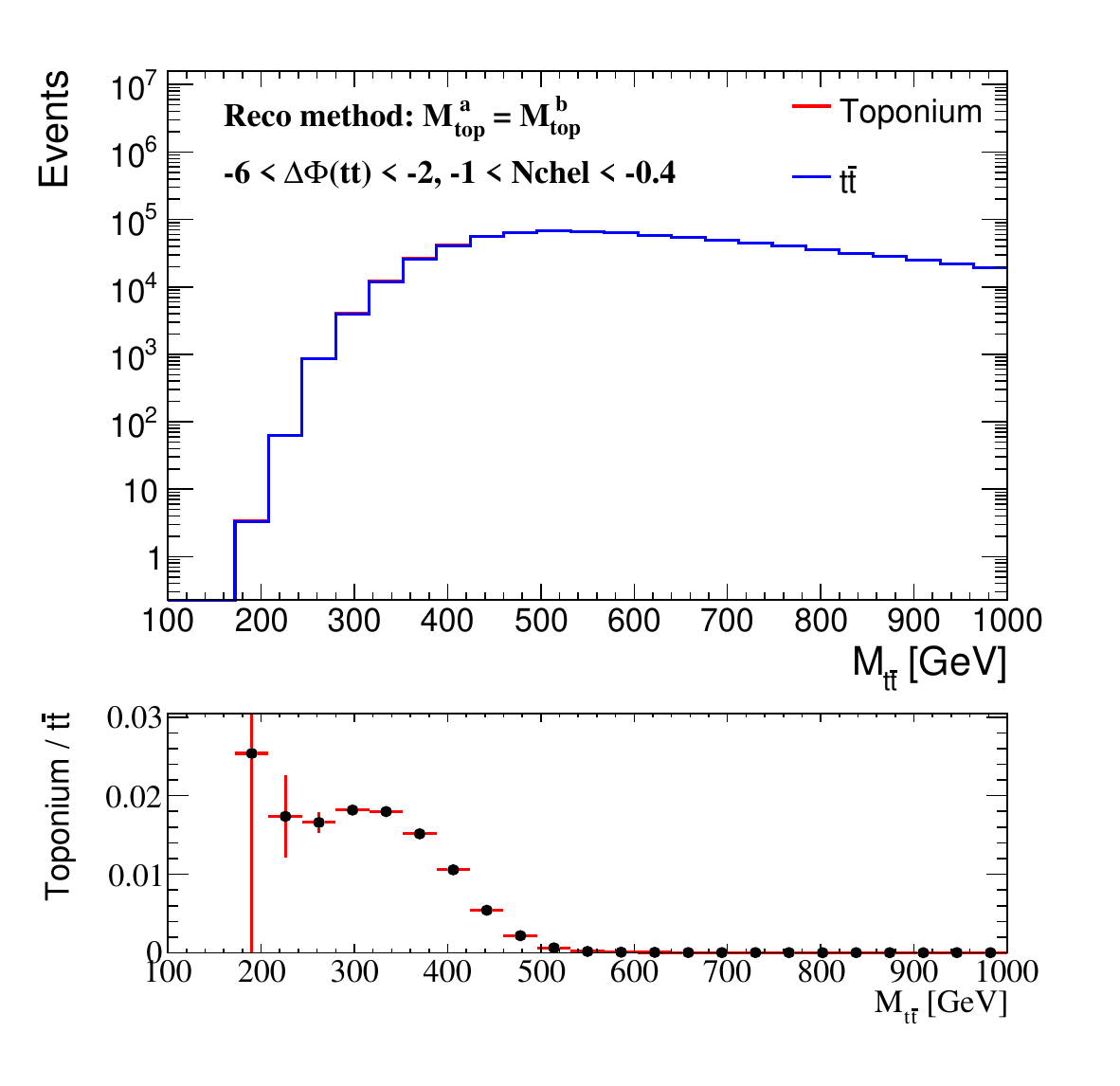}
        \caption{$(1,1)$}
        \label{fig:nchel_sig_11}
    \end{subfigure}
    \hfill
    \begin{subfigure}{0.32\textwidth}
        \centering
        \includegraphics[width=\linewidth]{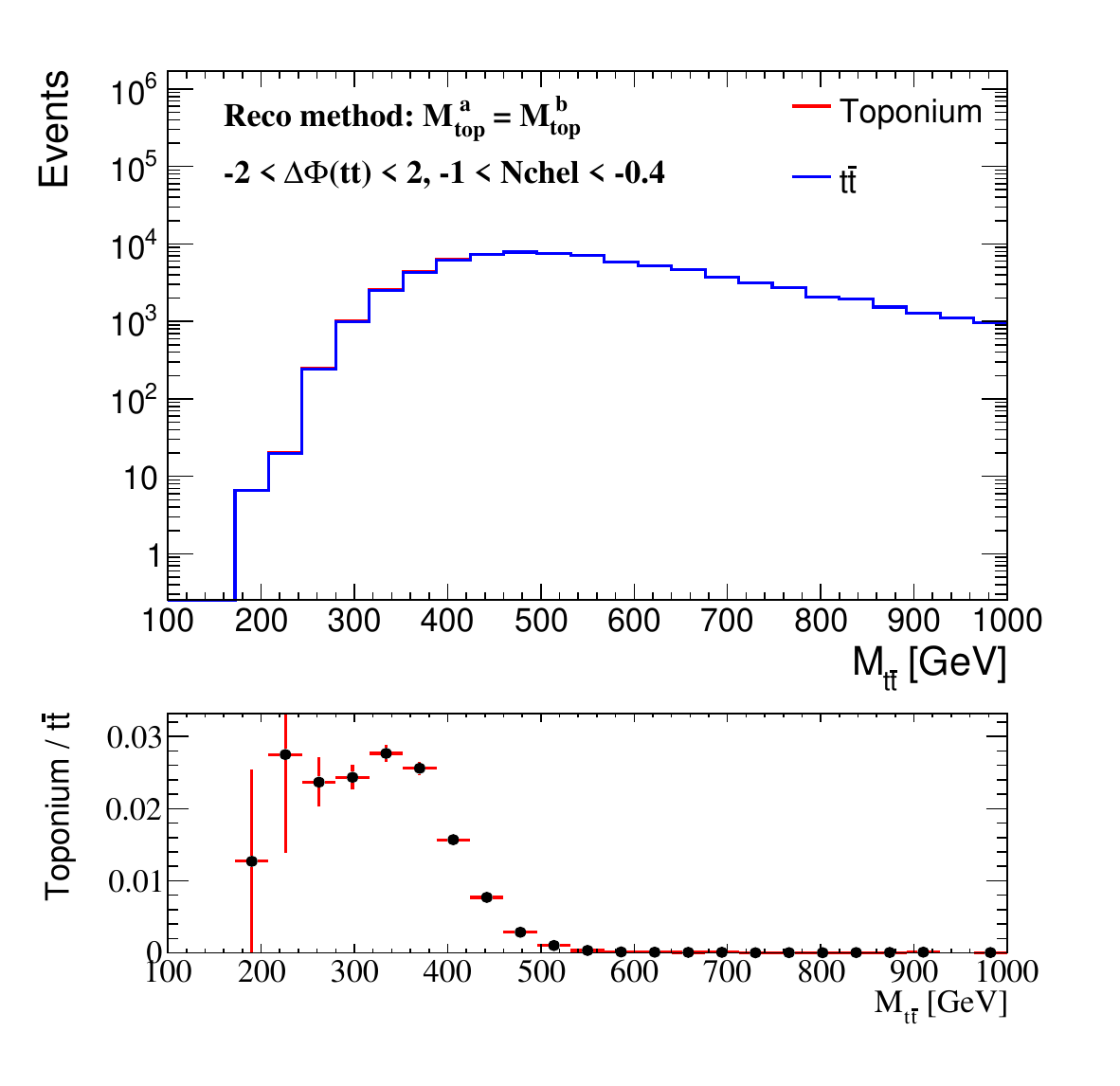}
        \caption{$(2,1)$}
        \label{fig:nchel_sig_21}
    \end{subfigure}
    \hfill
    \begin{subfigure}{0.32\textwidth}
        \centering
        \includegraphics[width=\linewidth]{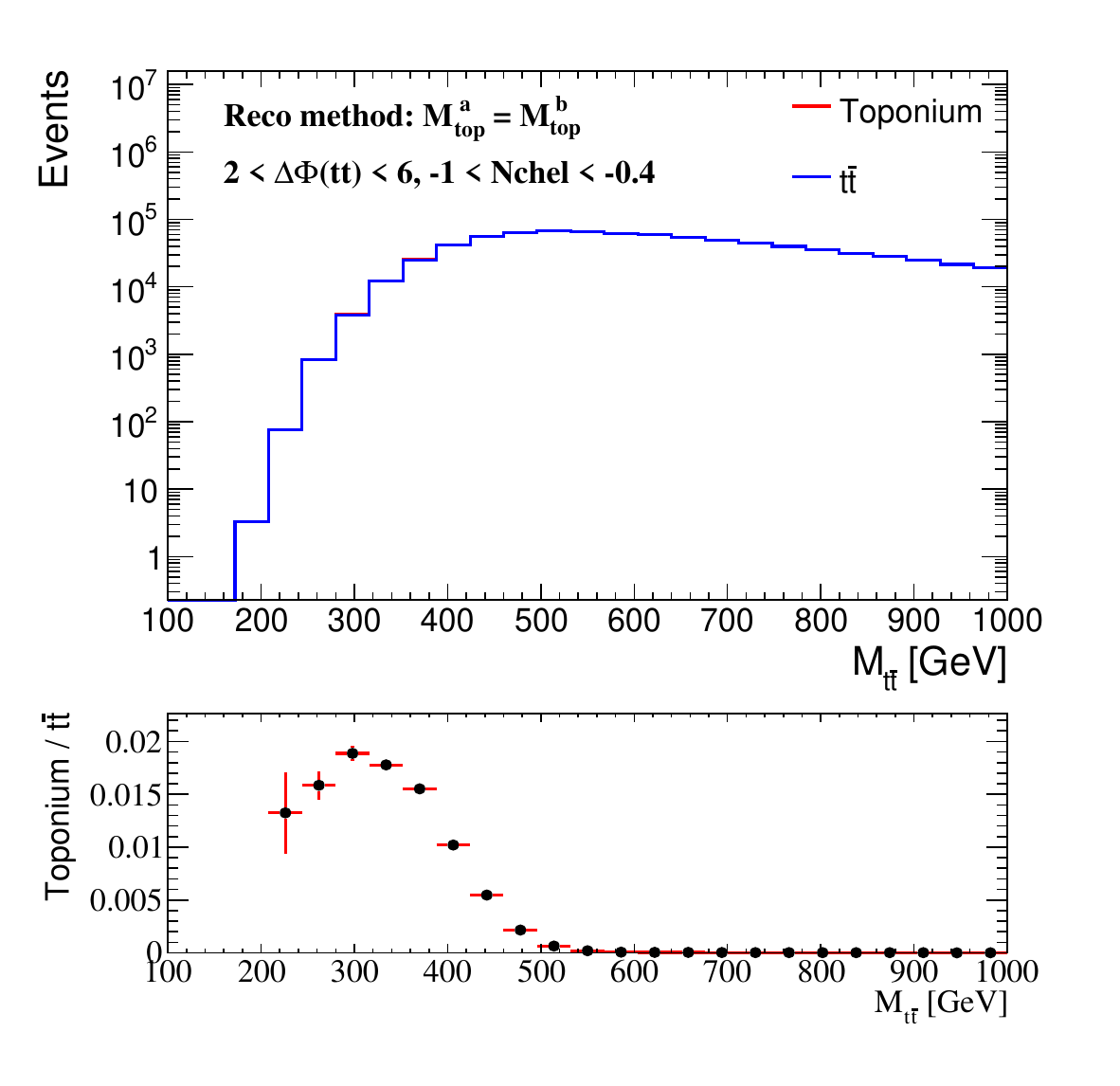}
        \caption{$(3,1)$}
        \label{fig:nchel_sig_31}
    \end{subfigure}

    \vspace{0.5em}

    \begin{subfigure}{0.32\textwidth}
        \centering
        \includegraphics[width=\linewidth]{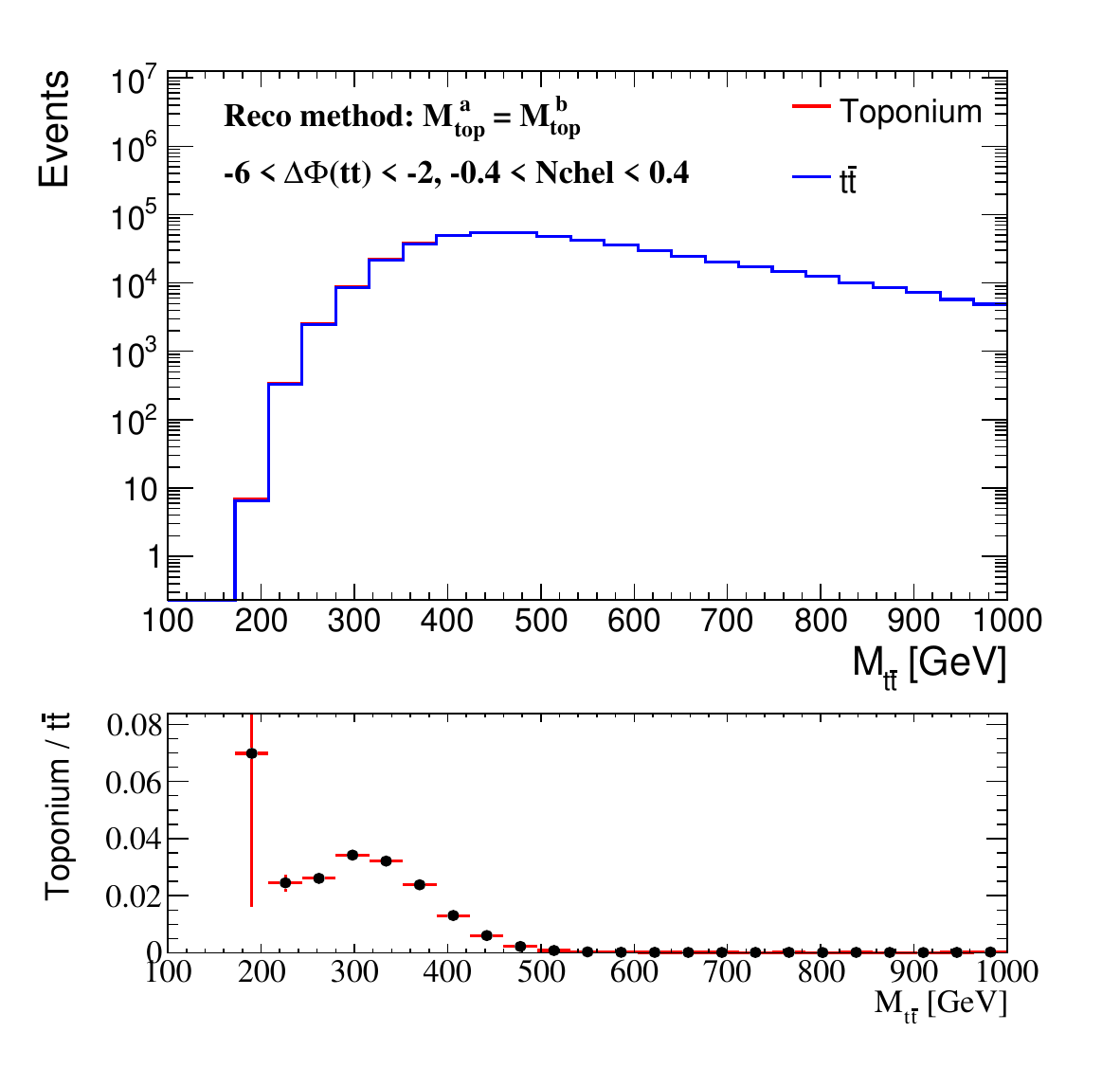}
        \caption{$(1,2)$}
        \label{fig:nchel_sig_12}
    \end{subfigure}
    \hfill
    \begin{subfigure}{0.32\textwidth}
        \centering
        \includegraphics[width=\linewidth]{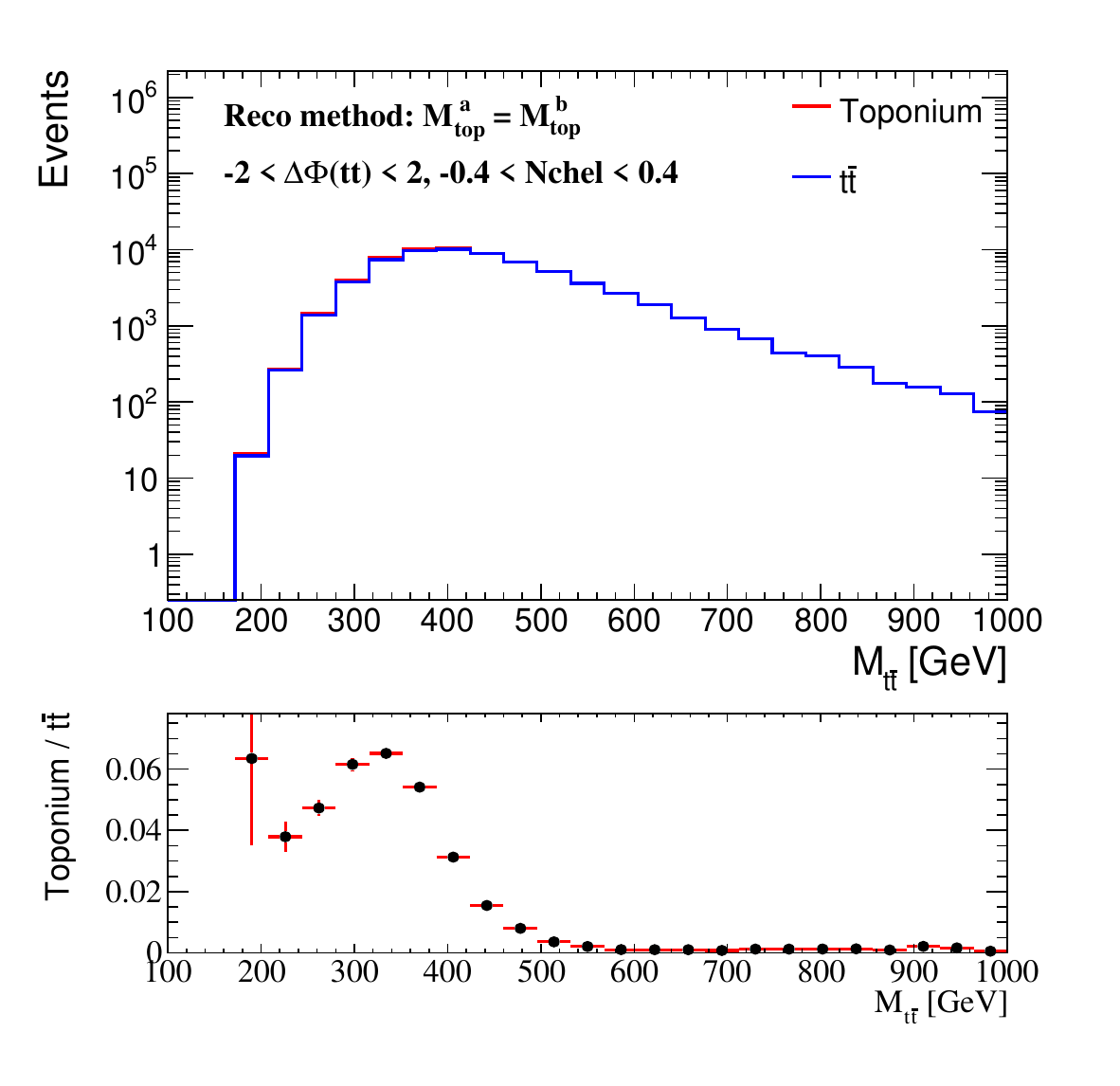}
        \caption{$(2,2)$}
        \label{fig:nchel_sig_22}
    \end{subfigure}
    \hfill
    \begin{subfigure}{0.32\textwidth}
        \centering
        \includegraphics[width=\linewidth]{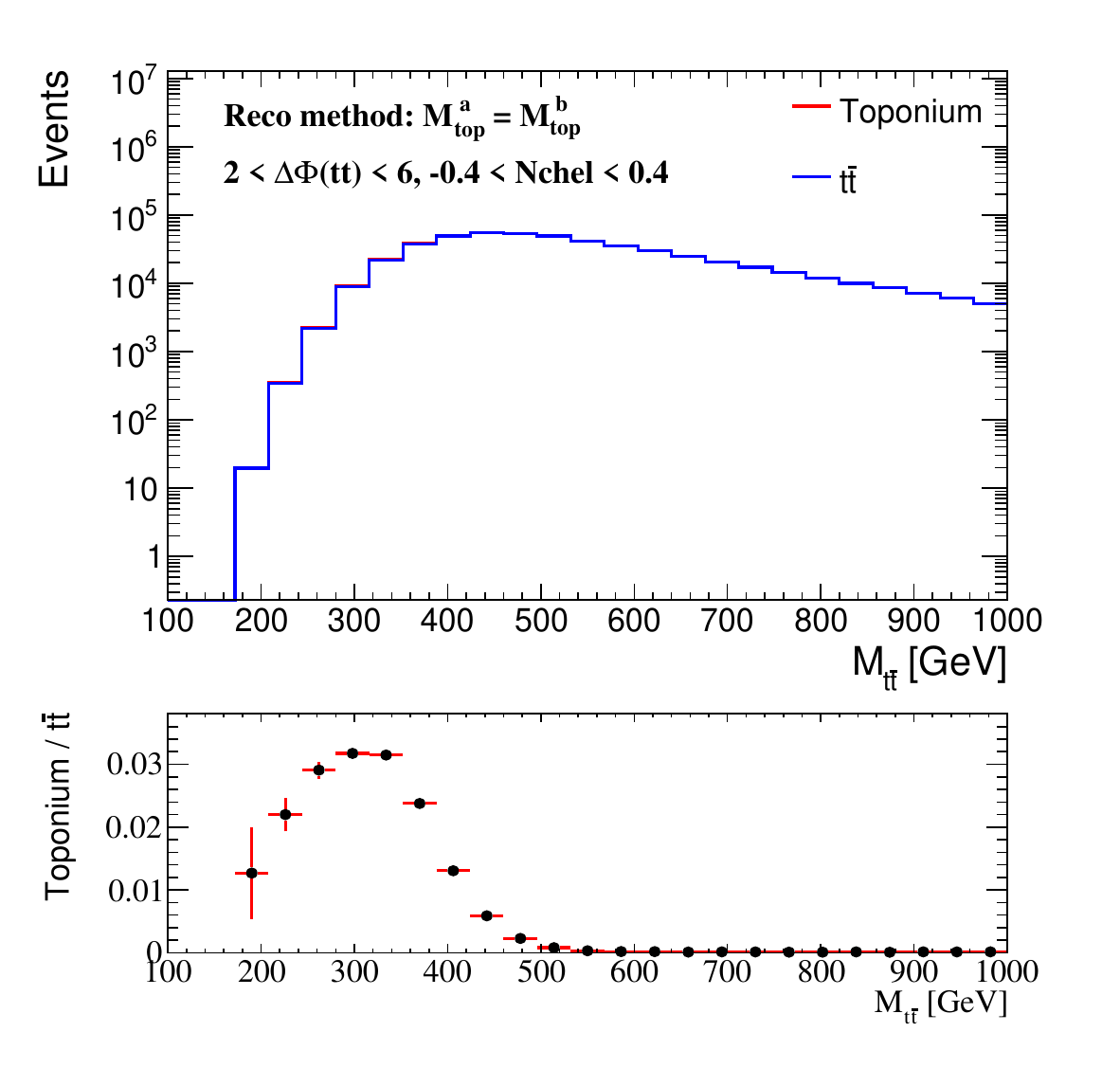}
        \caption{$(3,2)$}
        \label{fig:nchel_sig_32}
    \end{subfigure}

    \vspace{0.5em}

    \begin{subfigure}{0.32\textwidth}
        \centering
        \includegraphics[width=\linewidth]{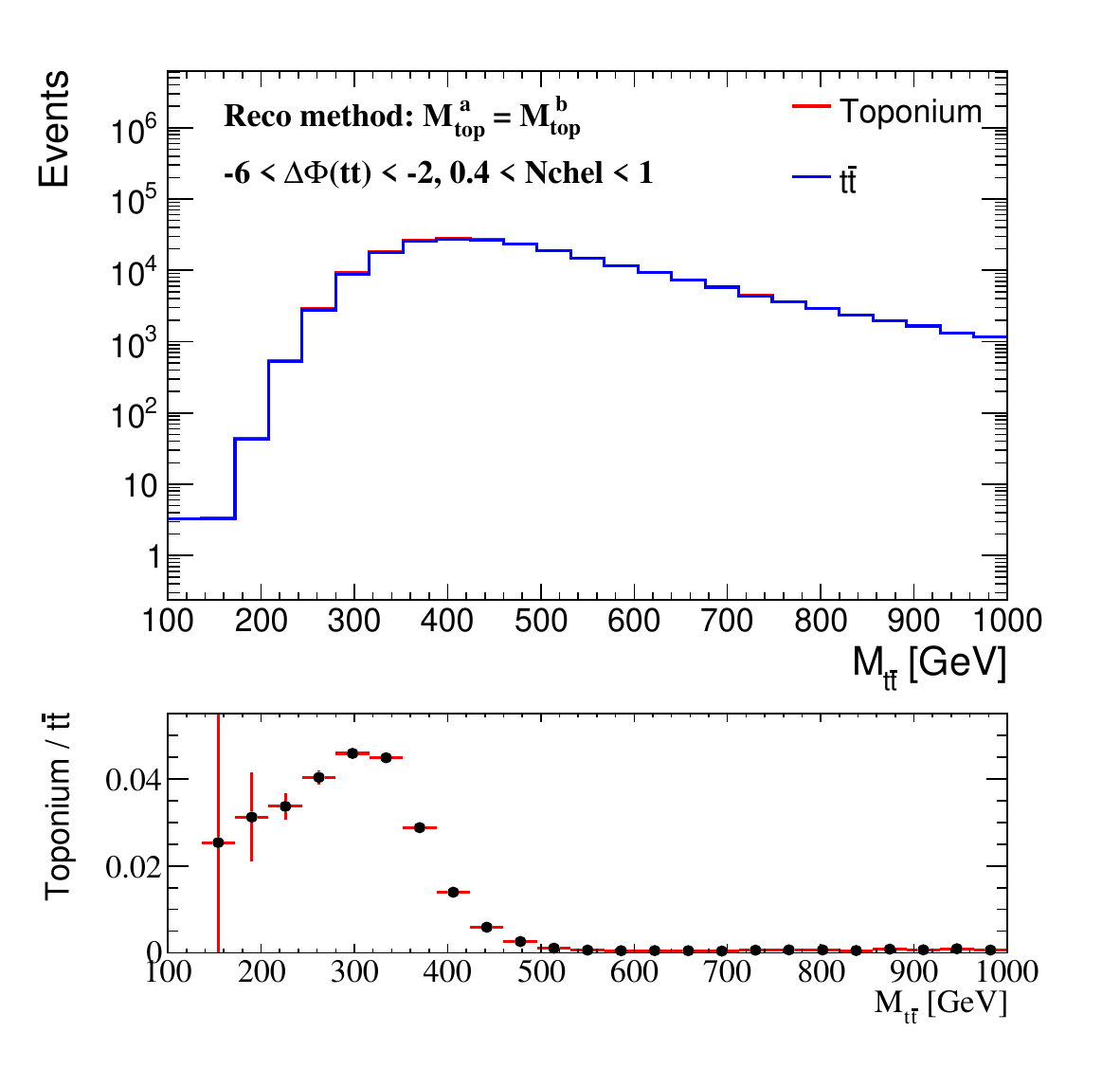}
        \caption{$(1,3)$}
        \label{fig:nchel_sig_13}
    \end{subfigure}
    \hfill
    \begin{subfigure}{0.32\textwidth}
        \centering
        \includegraphics[width=\linewidth]{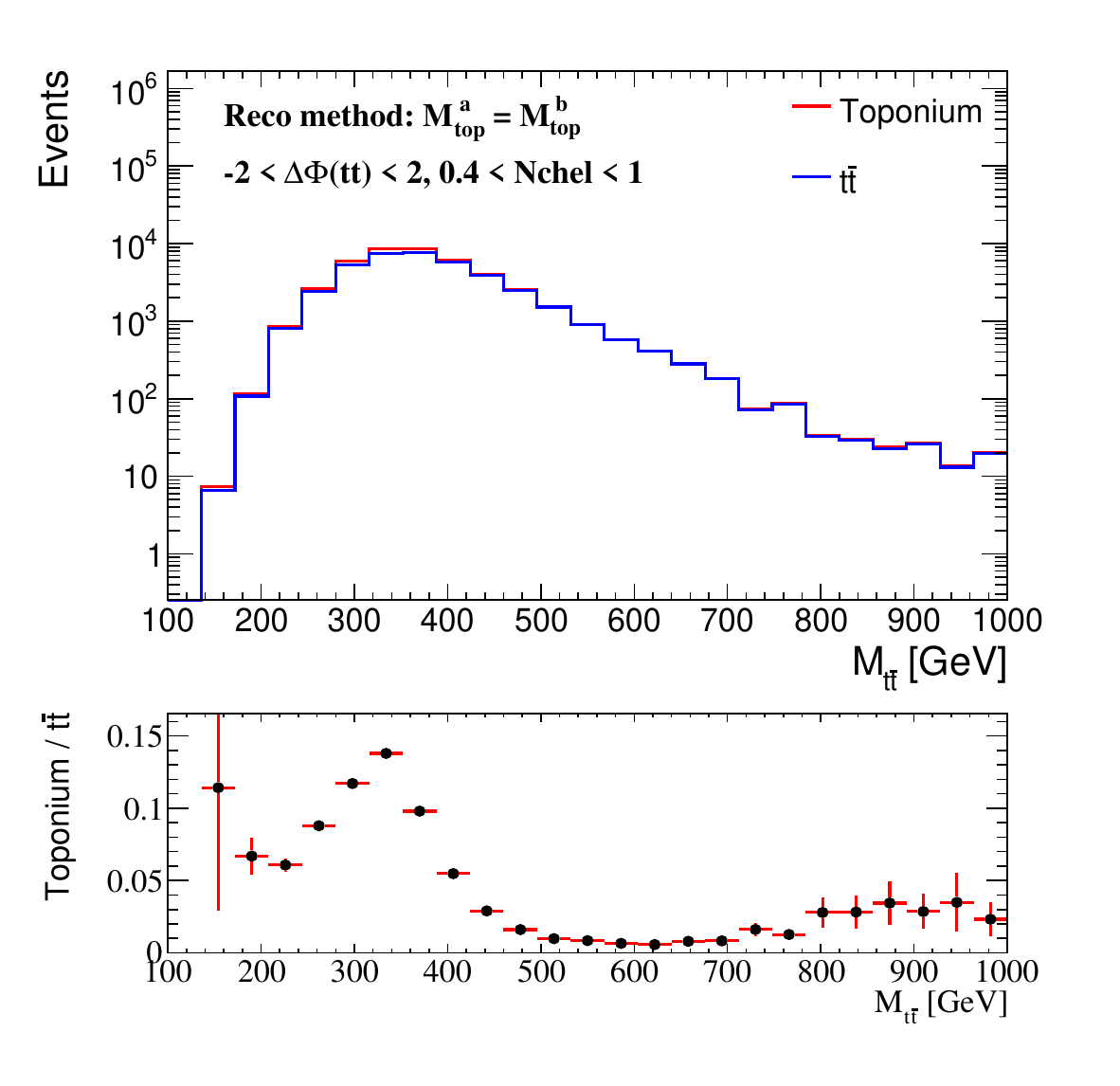}
        \caption{$(2,3)$}
        \label{fig:nchel_sig_23}
    \end{subfigure}
    \hfill
    \begin{subfigure}{0.32\textwidth}
        \centering
        \includegraphics[width=\linewidth]{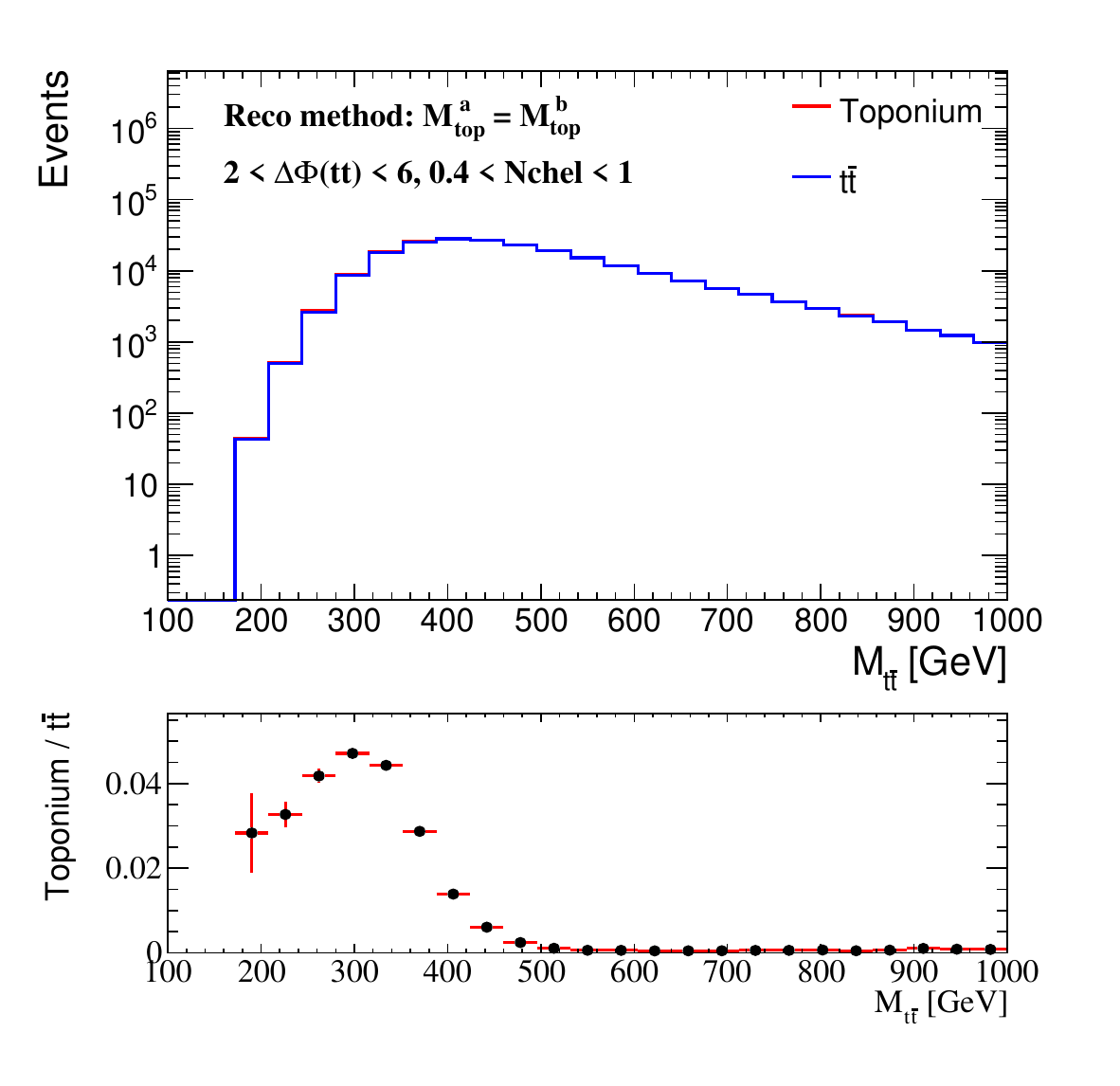}
        \caption{$(3,3)$}
        \label{fig:nchel_sig_33}
    \end{subfigure}

    \caption{Significance distributions for all $(i,j)$ bin combinations of $N_{\text{chel}}$ and $\Delta\Phi(t\bar{t})$, arranged column-wise following pre-selection criteria.}
    \label{fig:nchel_nine}
\end{figure*}

The $M_{\ttbar}$ distributions for all the nine regions are given in Figure~\ref{fig:nchel_nine}. In the plots, the optimal region with $\nchel \in (0.4,1) \otimes \Delta\Phi(\ttbar) \in (-2,2)$ has the significance of $15.5\sigma$. In the Run 2 configuration, in the optimal region $\nchel \in (0.4,1) \otimes \Delta\Phi(t,\bar{t}) \in (-2,2)$ one would see a significance of $10.2\sigma$ which is an improvement from using the $\chel \otimes \chan$ strategy.

On the other hand, one can also utilise the $\Delta\Phi(\ell,\ell)$ variable. One could utilise the following strategy: 

\begin{align}
\Delta\Phi(\ell,\ell) &\in \{[-6,-1.5],\, [-1.5,1.5],\, [1.5,6]\}, \\
\nchel &\in \{[-1,-0.4],\, [-0.4,0.4],\, [0.4,1]\}, \\[6pt]
\mathcal{R} &= \Delta\Phi(\ell,\ell) \text{bins} \otimes \nchel\text{bins}.
\end{align}

Following the strategy, the significance in the optimal region, $\nchel \in (0.4,1) \otimes \Delta\Phi(\ell,\ell) \in (-1.5,1.5)$,  is $14.9\sigma$ ($9.9\sigma$) for the Run 3 (Run 2) configuration.

\section{Analysis with Event Selection applied and evaluation of the Optimal Criteria}\label{sec:final}

We now consider applying a set of selection criteria to veto events that may arise from backgrounds other than $\ttbar$ following the experimental strategy. In particular, the following event selection is applied: Select events with $M(\ell\ell) > 15 ~\rm{GeV}$ to ensure that dileptons coming from low-mass resonance background are removed. Moreover, events are required to satisfy the condition that $|M_{\ell\ell} - M_Z| >  10 ~\rm{GeV}$ and, additionally, a condition on the missing transverse energy is imposed $E_T^{\mathrm{miss}} > 40~\rm{GeV}$. We also require $M_{\ttbar} < 550~\rm{GeV}$ to study the threshold region. The event selection criteria are summarised in Table~\ref{tab:event_selection}. 

Applying the new set of selection criteria, we reanalyze the phase space within the optimal region. In the $\chel\otimes\chan$ optimal region, one finds that the significance is $12.8\sigma$ in the Run 3 configuration. In the same way, for the optimal region of $\nchel  \otimes \Delta\Phi(t,\bar{t})$ the significance is found as $13.1\sigma$. The number is still an improvement over the $\chel\otimes\chan$ strategy.

\begin{table}[htbp]
\centering
\caption{Event selection criteria applied in the analysis.}
\label{tab:event_selection}
\begin{tabular}{@{} l c @{}}
\toprule
\textbf{Cut} & \textbf{Condition} \\
\midrule
Pre-selection & See Table~\ref{tab:preselection} \\
Dilepton invariant mass & $M(\ell\ell) > 15~\mathrm{GeV}$ \\
Z boson veto & $|M_{\ell\ell} - M_Z| > 10~\mathrm{GeV}$ \\
Missing transverse energy & $E_T^{\mathrm{miss}} > 40~\mathrm{GeV}$ \\
Top-quark pair invariant mass & $M_{\ttbar} < 550~\mathrm{GeV}$ \\
\bottomrule
\end{tabular}
\end{table}

In the Run 2 configuration, the significance in the  $\chel\otimes\chan$ optimal region is found to be $8.5\sigma$ whereas in the optimal region of $\nchel  \otimes \Delta\Phi(t,\bar{t})$ we find it as $8.6\sigma$.

\begin{figure}[tbp]
    \centering
    \includegraphics[width=0.7\linewidth]{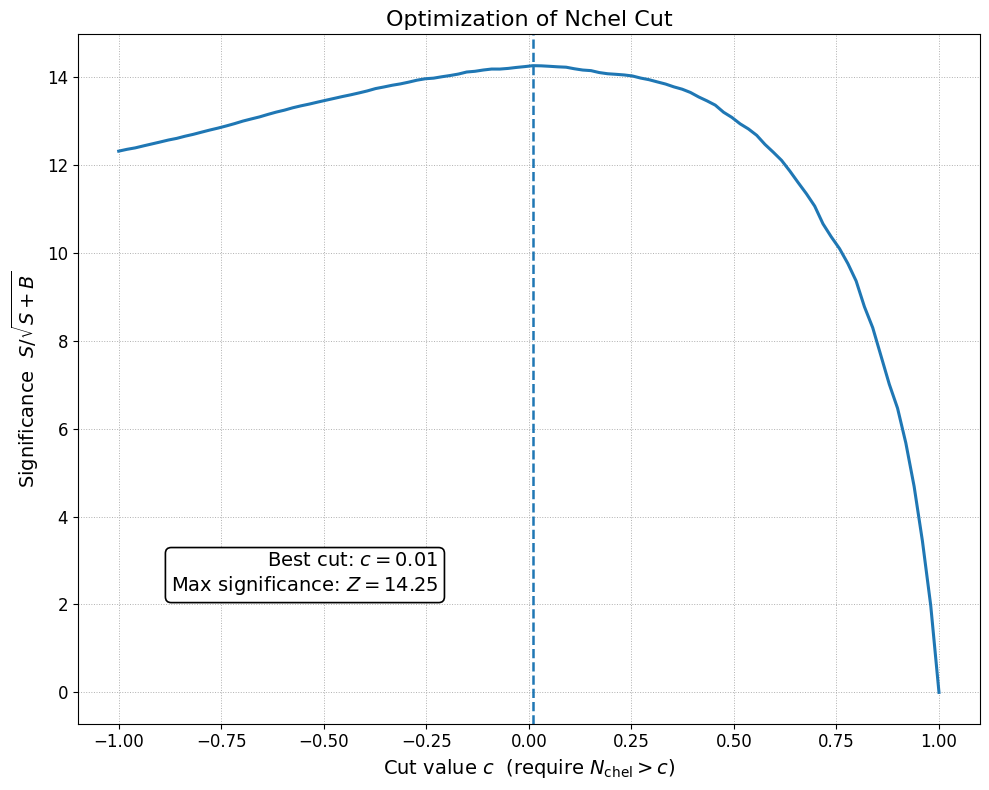}
    \caption{The optimal $\nchel$ cut applied after considering event selection criteria.}
    \label{fig:optimal_cut}
\end{figure}

However, note that the performance due to the new variable has only marginally improved the significance (in Run 3 configuration). In order to assess the actual impact due to the $\nchel$ variable, we optimise the $\nchel$ variable and the $\Delta\Phi(t,\bar{t})$ variable.  In particular, the evaluation is done by first choosing a smaller phase-space region for $|\Delta\Phi(\ttbar)| < 1.6$ whilst scanning the phase space for the optimal significance by applying selection criteria on $\nchel$ in $(-1,1)$. The result of this study is shown in Figure~\ref{fig:optimal_cut}. Here we see that the selection $\nchel>0.0$ will result in a significance of $14.3\sigma$ which is an improvement from the $\chel\otimes\chan$ optimal region.

\section{Conclusion}

We presented the Recursive Jigsaw Reconstruction method to reconstruct the toponium quasi-bound state at the Large Hadron Collider considering the LHC Run 2 and Run 3 collision energies. We propose a new analysis strategy where variables such as $\nchel$ and $\Delta\Phi(t,\bar{t})$ are introduced and which rely on the performance of the reconstruction framework. The $\nchel$ variable, which is the modified version of the $\chel$ variable, when combined with the $\Delta\Phi(t,\bar{t})$ has better discriminating power compared with the $\chel\otimes\chan$ strategy. Our results indicate that the maximum significance achieved using the $\chel\otimes\chan$ approach is about $12.4\sigma$ at LHC Run 3, and the newly proposed strategy ($\nchel  \otimes \Delta\Phi(t,\bar{t})$) can allow the significance as high as $15.5 \sigma$ before numerical optimisation applying only preselection requirements. On applying stricter event selection rules, one sees that the significance drops down. However, numerical optimisation on $\nchel$ results in a significance of $14.3\sigma$, which when compared to the $\chel\otimes\chan$ strategy leads to an improvement by 12\%. Furthermore, the Recursive Jigsaw Reconstruction provides additional scale sensitive and angular properties some of which were discussed in~\cite{Jackson:2017gcy}. These variables can be used to distinguish toponium from conventional $t\bar{t}$ and promise to improve discrimination between the two processes beyond the current level. Such variables will also enhance sensitivity to the search for evidence of the existence of additional states.


\bibliographystyle{apsrev4-2}
\bibliography{refs}

\end{document}